\documentclass[10pt,onecolumn,peerreview]{IEEEtran}
\usepackage{makeidx}
\usepackage[utf8]{inputenc}
\usepackage[dvips]{graphicx}
\usepackage{graphicx,url}
\usepackage{amssymb}%fancyhdr,tabularx,epsfig}
\usepackage{psfrag,amsmath}
\usepackage{multirow}
\usepackage[Lenny]{fncychap}
\usepackage{fancyhdr}
\usepackage{makeidx}
\usepackage{epsfig}
\usepackage{graphicx,color}
\usepackage{subfig}
\usepackage{array}
\usepackage{arydshln}
\usepackage[all]{xy}
\usepackage{float}
\usepackage{hyperref}
\usepackage{multicol}

\usepackage[T1]{fontenc}
\usepackage{ifthen}
\usepackage{cite}
\usepackage{stfloats} %pacote para colocar a tabela/figura em qualquer parte de um texto de duas colunas 
\usepackage{blindtext} %tabelas com duas colunas de texto 
\newtheorem{coro}{Corollary}

\newtheorem{defi}{Definition}
\newtheorem{teo}{Theorem}
\newtheorem{lema}{Lemma}
\newtheorem{exem}{Example}
\newtheorem{prop}{Proposition}
\newtheorem{rema}{Remark}
\newcommand{\Fn}{\mathbb{F}_2^n}
\newcommand{\Fm}{\mathbb{F}_2^m}
\newcommand{\Fnm}{\mathbb{F}_2^{n+m}}
\newcommand{\Fmn}{\mathbb{F}_2^{m+n}}
\newcommand{\FF}{\mathcal{F}}
\newcommand{\supp}{\mathrm{supp}}
\newcommand{\argmin}{\mathrm{arg \ min}}
\newcommand{\bx}{\mathbf{x}}
\newcommand{\by}{\mathbf{y}}
\newcommand{\bz}{\mathbf{z}}
\newcommand{\bzero}{\mathbf{0}}
\newcommand{\bc}{\mathbf{c}}
\newcommand{\be}{\mathbf{e}}
\newcommand{\ba}{\mathbf{a}}
\newcommand{\bb}{\mathbf{b}}
\newcommand{\bw}{\mathbf{w}}
\newcommand{\bv}{\mathbf{v}}
\newcommand{\rank}{\mathrm{rank}}

\usepackage{setspace}

\begin{document}
\onehalfspacing
		\title{Obtaining binary perfect codes out of tilings} 
	% %%% Single author, or several authors with same affiliation:
	% \author{%
	%   \IEEEauthorblockN{Stefan M.~Moser}
	%   \IEEEauthorblockA{ETH Zürich\\
	%                     ISI (D-ITET)\\
	%                     CH-8092 Zürich, Switzerland\\
	%                     Email: moser@isi.ee.ethz.ch}
	% }

	%%% Several authors with up to three affiliations:
\author{Gabriella Akemi Miyamoto \thanks{A reduced version of this paper was presented at ISIT  2019.}
 and Marcelo Firer} 

\maketitle

\begin{abstract}

A tiling of the $n$-dimensional Hamming cube gives rise to a perfect code (according to a given metric) if the basic tile is a metric ball. We are concerned with metrics on the $n$-dimensional Hamming cube which are determined by a weight which respects support of vectors (TS-metrics). We consider the known tilings of the Hamming cube and first determine which of them give rise to a perfect code. In the sequence, for those tilings that satisfy this condition, we determine all the TS-metrics that turns it into a perfect code. We also propose the construction of new perfect codes obtained by the concatenation of two smaller ones.
\end{abstract}

\section{Introduction}

Perfect codes have been extensively studied in the literature, due to the optimality condition imbued in its definition and to the interesting challenges they pose. 
Nevertheless, considering the Hamming metric, perfect codes are rare and they are classified in the case of linear binary codes: It has the same parameters of binary repetition codes with odd length, $q$-ary Hamming codes, binary or ternary Golay codes, \cite{hamming,golay1}. The situation is not the same for the Lee metric. There are few results (see \cite{golomb2, golomb,post}), but there is not a complete characterization. Two good surveys about perfect codes in Hamming metric are \cite{lint,olof}. 

\vspace{3pt}

The most primary definition of a perfect code is the geometrical one: a code is perfect if its packing radius equals its covering radius. This definition is interesting in our setting since it depends directly on the metric invariants, which are naturally defined in general settings. To be more explicit, given a metric space $(X,d)$ and a subset $C\subset X$, we define its \textbf{packing radius} $R_{d,pack}(C)$ as the maximal $r$ such the balls of radius $r$ centered at elements of $C$ are disjoint and its \textbf{covering radius} $R_{d,cov}(C)$ as the minimum $r$ such that the balls of radius $r$ centered at elements of $C$ covers the space $X$. The set $C$ is called a $d$-perfect code if $R_{d,pack}(C)=R_{d,cov}(C)$. Set this, it is understandable that the study of perfect codes can be undertaken considering a more vast family of metrics and it can be valuable to do so for any metric that has some relevance in the context of coding theory.

\vspace{3pt}

Considering this scarcity of perfect codes under the Hamming metric, the introduction of the poset metrics by Brualdi et al. in 1995 \cite{brualdi} drawn the attention since, in general, there is a relative abundance of perfect codes (depending on the poset metric). The study of perfect codes in this context is  done in different approaches. One of them is to fix a particular family of posets (chain, crown or hierarchical) and to classify all the perfect codes for the particular family, as done in \cite{jang2,kim,panek}. Another approach is to  consider a family of well known codes and asks to classify all the poset metrics which turn the code to be perfect. This is what is done, for example, with the extended Hamming and Golay codes for poset metrics (in \cite{hyun,jang}) and for poset-block metrics (\cite{muniz,dass}). Our approach resembles more the second one, but instead of looking at the codes, we fix the tiles.

%At this point we should remark that the relative abundance of perfect codes with a poset metric is one of the features that first attracted the attention to this family of metric. In the literature we can find two different approaches to this question. The first one  Another approach is to 
%
%The most common approach used consists of fixing a metric (either the Hamming  or a poset metric) and looking for perfect codes (BUSCAR REFERENCIAS) or of obstructions for its existence, such as determining the parameters of a perfect code (BUSCAR REFERENCIAS). 

We wish to find new families of perfect codes, but our approach is somehow different. We consider some configurations of points in the $n$-dimensional Hamming cube $\mathcal{H}_n$ (called tilings) defined by some properties that are similar to the ones that defines a perfect code in a vector space, except for one point: the tiles may not be metric balls. A priori, we fix no metric, but rather look for metrics which turns the tiles into a metric ball and we get a perfect code for this metric.

We do not consider the whole universe of metrics that can be defined on $\mathcal{H}_n$, but restrict ourselves to a (vast) family of metrics, which we call TS-metrics. The \textbf{T} stands for a metric that is invariant by \textbf{t}ranslations, or equivalently, determined by a weight (as the Hamming metric is determined by the Hamming weight and vice-versa). The \textbf{S} stands for a metric that respects the \textbf{s}upport, in the sense that if $\bx,\by\in\Fn$ are vectors such that the $i$-th coordinate of $\bx$ is nonzero implies that the  $i$-th coordinate of $\by$ is also nonzero, then the weight of $\bx$ is not greater than the weight of $\by$.

% resembles the  to find new perfect codes can be considered as the reverse of the conventional one. Given a tile $(D,C)$ of $\Fn$, we try to find some TS-metric such that $D$ is a ball and then, $C$ is a perfect code.

Among the TS-metrics, there are two important and large  families: the poset metrics (introduced in \cite{brualdi}) and the combinatorial metrics (introduced in \cite{gabidulin}). 

%The poset metrics were  and there is some results about perfect codes in these metrics. The combinatorial metrics were presented in \cite{gabidulin}. The study of perfect codes under the poset metrics has been extensively studied (see REFERENCIAS DA SESSÃO 6.3.1) since, as noted already by Brualdi et al. \cite{brualdi}, they are more abundant for general poset metrics than in the case of the Hamming metric.

Our main working scheme consists of the following steps: 
\begin{enumerate}
	\item Look for a tiling $(D,C)$ of the Hamming cube $\Fn$; 
	\item Consider the tile $D$ and discharge all those ones that can not be a ball by considering any TS-metric;
	\item For those that can not be discharged we should look for a metric that turns it into a ball; 
	\item We try to classify (in a sense that will be explained later) all such metrics; 
	\item Finally, considering the tiles that are balls, we consider them as a kind of basic bricks and try to determine some ways we can use them to build ``larger'' perfect codes.
\end{enumerate}

The starting point of this script is to find tilings of the Hamming cube, either considering as a vector space or as a graph. Not much is known on the subject. Despite the existence of some very constructions (\cite{gruslys,gruslys2}), these constructions are discharged at the second step, so that it does not push over our purpose. So, our source to start the procedure is essentially the remarkable paper \cite{vardy} by Cohen, Lystin, Vardy and Z\'{e}mor.  In this work the authors characterized all the 193 tiles with 8 elements and also tilings where the tiles have high rank (see details in Section \ref{sec:concatenation}). This will be our starting point.

This work is organized as follows: In Section \ref{sec:prelimi} we introduce all the preliminary definitions and notations, about TS-metrics \ref{sec:metrics}, including poset and combinatorial metrics, perfect codes \ref{sec:perfect codes}, tiles and tilings \ref{sec:tilings} and concatenation of tilings \ref{sec:concatenation}; in Section \ref{sec:obtain} we develop the first four steps of our program, that is, how to obtain perfect codes out of tilings; finally, the last step, that is, how to build new perfect codes out of known one is studied in Section \ref{sec:concatenation and perfect codes}.
  
%In this work, we present the small tiles of \cite{vardy} and show which one is a ball for some TS-metric, obtaining a perfect code. After that, we establish which are all the metrics that turns the tile into a ball. Finally, we show which are the TS-metrics that turns a concatenation of perfect codes into a perfect code.   

\section{Preliminaries}\label{sec:prelimi}

Let $\mathbb{F}_2^n$ be the $n$-dimensional vector space over $\mathbb{F}_2$ and let $\omega_H$ and $d_H$ denote the Hamming weight and the Hamming metric, respectively.  The \textbf{support} of a vector $\bx\in \mathbb{F}_2^n$ is  the set $\supp(\bx):=\{i\in [n] : x_i\not= 0\}$, where $[n]=\{1,2,\ldots,n\}$. 

\vspace{3pt}

In this section we introduce the basic concepts, definitions and notations used in this work: the translation-invariant and respecting support metrics (TS-metrics) together with the two best known  families of TS-metrics (poset and combinatorial metrics) in Section \ref{sec:metrics}; perfect codes and TS-perfect codes (Section \ref{sec:perfect codes}); tilings of $\mathbb{F}_2^n$ and polyhedrominoes (Section \ref{sec:tilings}) and  concatenation of tilings  (Section \ref{sec:concatenation}).

% conditional sums of metrics (Sections \ref{extensaotile}) and \ref{sec:concatenation and perfect codes}.       

\subsection{TS-metrics}\label{sec:metrics}

The Hamming metric has two important properties, expressed in the next two definitions.
\vspace{3pt}

\begin{defi}
	A metric $d:\mathbb{F}_2^n \times \mathbb{F}_2^n \to \mathbb{R}$ is said to be \textbf{translation-invariant} if 
	\[d(\bx+\bz,\by+\bz)=d(\bx,\by)\]
	for every $\bx ,\by,\bz\in \mathbb{F}_2^n$.
\end{defi}

It is well known and worth noting that a metric is translation-invariant \emph{iff} it is defined by a weight \footnote{A function $\omega:\mathbb{F}_2^n\to \mathbb{R}$ is a \textbf{weight} if it satisfies the following axioms: $(1)$  $\omega(\bx)\geq 0$ for every $\bx$; $(2)$  $\omega(\bx)=0$ \textit{iff} $\bx=0$; $(3)$ $\omega(\bx+\by)\leq \omega(\bx)+\omega(\by)$. A weight determines a metric by defining $d(\bx,\by)=\omega(\bx-\by)$. }.

\vspace{3pt}
\begin{defi}
	A weight function $\omega$ \textbf{respects the support} of vectors if 
	$\supp(\bx)\subseteq \supp(\by) \Longrightarrow \omega(\bx)\leq \omega(\by).$
\end{defi}

\vspace{3pt}
	A \textbf{translation-support metric} (\textbf{TS-metric}) is a metric which satisfies both the properties, that is, it is translation-invariant and it respects the support of vectors.
	
	In this work we restrict ourselves to TS-metrics, a restriction that is reasonable because: 1) being translation-invariant is a key property for decoding linear codes, since syndrome decoding depends exclusively on this property; 2) respecting the support of vectors is a property that is crucial in coding theory (for binary codes), once it means that making extra errors cannot lead to a better situation, in the sense that making an error on the $i$-th coordinate of a message cannot be worse than making two errors, one on the $i$-th coordinate  and the other on the $j$-th. %By the translation-invariant property, it is enough to take $0$ as the center of the balls. 

\vspace{3pt}

The set ${\mathcal{TS}(n)}$ of all possible TS-metrics on $\mathbb{F}_2^n$ is not well described or studied, but it contains two well understood large families that can be used as \emph{bricks} out of which every TS-metric can be build (for details see \cite{roberto}). Those are the families of poset and combinatorial metrics, which we now introduce. 
\vspace{3pt}

\subsubsection{Poset metric}

In its full generality, the poset metrics were introduced by Brualdi et al. in \cite{brualdi}. For more details, see a recent survey on \cite{marcelo}.

Let $P=([n],\preceq)$ be a partially ordered set (\textbf{poset}).  An \textbf{ideal} in a poset $P=([n],\preceq)$ is a nonempty subset $I\subseteq [n]$ such that, for $a\in I$ and $b\in [n]$, if $b\preceq a$ then $b\in I$. We denote by $\langle A\rangle$ the ideal generated by $A\subseteq [n]$. 

An element $a$ of an ideal $I\subset [n]$ is called a \textbf{maximal} element of $I$ if $a\preceq b$ for some $b\in I$ implies $b=a$. %The set of all maximal elements of an ideal $I$ is denoted by $\mathcal{M}_{P}(I)$. Notice that, given an ideal $I\subseteq [n]$, we have that $\langle \mathcal{M}_{P}(I) \rangle_{P}=I$ and $I$ is minimal with this property. 
 
We say that $b$ \textbf{covers} $a$ if $a\preceq b$, $a\not= b$ and there is no extra element $c\in [n]$ such that $a\preceq c\preceq b$. If $b$ covers $a$, then $(a,b)$ is said the \textbf{covering pair}. 
\vspace{3pt}

\begin{defi}
The \textbf{$P$-weight} of a vector $\bx\in\mathbb{F}_2^n$ is defined by
\[\omega_{P}(\bx):=|\langle \supp(\bx)\rangle|,\]
where $|A|$ is the cardinality of $A$.
\end{defi}
\vspace{3pt}

The $P$-weight clearly respects support, since $A\subset B$ implies $\langle A\rangle \subset \langle B \rangle$. The \textbf{$P$-distance} in $\mathbb{F}_2^n$ is the metric induced by $\omega_{P}$: $d_{P}(\bx,\by):=\omega_{P}(\bx-\by)$.

\vspace{3pt}

It is possible to geometrically describe a poset using the so called Hasse diagram. The \textbf{Hasse diagram} of a poset $P=([n],\preceq)$ is the directed graph with vertex set $[n]$ and whose edges are the covering pairs $(\bx,\by)$ in $P$. The Hasse diagram is picturing assuming that, given a covering pair $(\bx,\by)$ with $\bx\preceq\by$ then $\by$ is ``above'' $\bx$.

\vspace{3pt}

\begin{exem}
	Consider $[5]=\{1,2,3,4,5\}$ and the poset $P: 1\preceq 4, 2\preceq 4, 3\preceq 4, 3\preceq 5$. The Hasse diagram of $P$ is given by 
	
	\begin{figure}[H]
		\begin{center}
			\includegraphics[width=2.3cm]{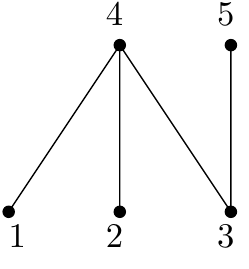}
			\caption{Hasse diagram of $P=([5],\preceq)$.}
		\end{center}
	\end{figure}
\end{exem}
 
\subsubsection{Combinatorial metric} 

The combinatorial metrics were introduced by Gabidulin in \cite{gabidulin}. For more details see \cite{bossert,feng,jerry}.

Let $\mathbb{P}_n=\{A; A\subset[n]\}$ be the power set of $[n]$. We say that a family $\mathcal{A}\subset \mathbb{P}_n$ is a \textbf{covering} of a set $X\subset[n]$ if $X\subset \displaystyle\bigcup_{A\in\mathcal{A}}A$. 

If $\mathcal{F}$ is a covering of $[n]$, then the \textbf{$\mathcal{F}$-combinatorial weight} of $\bx=(x_1,\ldots,x_n)\in \mathbb{F}_2^n$ is the integer-valued map $\omega_{\mathcal{F}}$ defined by
\[\omega_{\mathcal{F}}(\bx):=\mbox{min}\{|\mathcal{A}| ; \mathcal{A}\subset \mathcal{F}, \mathcal{A} \text{ is a  covering of } \supp(\bx) \}.\]

The function $d_{\mathcal{F}}(\bx,\by):=\omega_{\mathcal{F}}(\bx-\by)$ is a distance, which defines the \textbf{$\mathcal{F}$-combinatorial metric}.

We denote by  ${\mathcal{P}(n)}$ and ${\mathcal{C}(n)}$, respectively, the sets of all  poset and combinatorial metrics on $\mathbb{F}_2^n$. It is worth to noting that $\mathcal{P}(n)$ and $\mathcal{C}(n)$ are both subsets of $\mathcal{TS}(n)$.  

\subsection{Perfect codes}\label{sec:perfect codes}
Given a metric  $d$ on $\mathbb{F}_2^n$, the \textbf{ball of radius $r$ and center $\bx$} is 
		$B_d(\bx,r):=\{ \by\in \mathbb{F}_2^n; d(\bx,\by)\leq r\}.$ A code $C\subseteq \mathbb{F}_2^n$ is a $\mathbf{(d,r)}$\textbf{-perfect code} if its covering radius coincides with its packing radius, that is, 
		\[\displaystyle \bigcup_{\bc\in C} B_d(\bc,r) =\mathbb{F}_2^n \ \
		\mbox{and} \ \
		B_d(\bc,r) \cap B_d(\bc',r) =\emptyset, \forall \bc,\bc'\in C, \bc\neq\bc'.\]
	
We approach now the first of our key definitions. 
\vspace{3pt}

\begin{defi}
Given a subset $S\subseteq \Fn$, we say that $S$ is a \textbf{TS-ball} if $S$ is a ball for some TS-metric, that is, $S=B_d(\bx,r)$, for some $\bx\in \Fn$, $r>0$ and some $d\in \mathcal{TS}(n)$. If $C$ is a $(d,r)$-perfect code for some $d\in\mathcal{TS}(n)$ we say that it is a \textbf{TS-perfect code}. In case the radius $r$ is not taken into consideration, we say that $C$ is $d$-perfect.
\end{defi}

\vspace{3pt}

\begin{exem}
	Let $D=\{0000,1000,0100,0010,1100,1010\}$ be a set. Consider the poset $P_1$ represented by the Hasse diagram 
	\begin{figure}[H]
		\begin{center}
			\includegraphics[width=1.5cm]{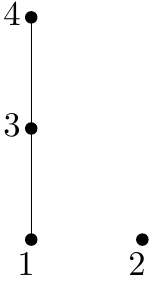}
			\caption{Hasse diagram of $P_1=([4],\preceq)$}
		\end{center}
	\end{figure}	

Then, we have $D=B_{d_{P_1}}(\bzero,2)$. Notice that $P_1$ is not the only poset that turns $D$ into a ball. If we consider $P_2$ given by the Hasse diagram in Figure \ref{N}, we would also have that $D=B_{d_{P_2}}(\bzero,2)$.
	
		\begin{figure}[H]
		\begin{center}
			\includegraphics[width=1.8cm]{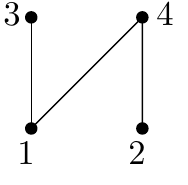}
			\caption{Hasse diagram of $P_2=([4],\preceq)$}
			\label{N}
		\end{center}
	\end{figure}

Now, consider \[D=\{0000,1000,0100,0010,0001,1100,1010\}.\] 

Given $\FF=\{\{1,2\},\{1,3\},\{4\}\}$ we have that $D=B_{d_{\FF}}(\bzero, 1)$. 
\end{exem}

\subsection{Tiles, tilings  and polyhedrominoes}\label{sec:tilings}

%\subsection{Tilings of $\mathbb{F}_2^n$}

We are interested in building perfect codes out of tilings of the Hamming cube, so we need some basic definitions about tilings and polyhedrominoes.

 A \textbf{path}  in $\mathbb{F}_2^n$, with initial point $\bx$ and final point $\by$, is a sequence of points $\gamma:\bx_{0},\bx_{1},\ldots,\bx_{t}$, where $d_{H}\left(  \bx_{i},\bx_{i+1}\right)  =1$, $\bx=\bx_{0}$ and $\by=\bx_{t}$. The \textbf{length} of $\gamma$ is defined by $\left\vert \gamma\right\vert =t $. A path $\gamma$ is called a \textbf{geodesic path} if it is a path of minimum length between the initial and final points.  It is easy to see that a path $\gamma$ from $\bx$ to $\by$ is  geodesic \emph{iff} $d_{H}\left(  \bx,\by\right)=\left\vert \gamma\right\vert $.

%A subset $V\subset\mathbb{F}_{2}^{n}$ is said to be \textbf{connected} if given $v,w\in V$ exists a path $\gamma$ connecting $v$ to $w$ with $\gamma\subset V$. $V$ is a \textbf{convex set} if exists a geodesic $\gamma$ connecting $v$ to $w$ with $\gamma\subset V$. Obviously, any convex set is also connected, but the converse is not true.
%
%A subset $V\subset\mathbb{F}_{2}^{n}$ is \textbf{strongly convex at a point $x\in V$}  if given $v\in V$, \textbf{every} geodesic $\gamma$ connecting $x$ to $v$ satisfies $\gamma\subset V$.

% If $D$ is a strongly convex set related to the center $x$, then $D$ is a \textbf{convex polyhedromino}.
\vspace{3pt}

\begin{defi}
A set $D\subseteq \mathbb{F}_2^n$ is a \textbf{polyhedromino} if for all $\bx,\by\in D$ there is a (possibly not unique) geodesic path $\gamma\subset D$ connecting $\bx$ to $\by$. We say that $D$ is a \textbf{convex polyhedromino} if, for every $\bx,\by \in D$, every geodesic path $\gamma$ connecting $\bx$ to $\by$ is contained in $D$. 
	
\end{defi}
\vspace{3pt}

The concept of tiling is used in many different continuous and discrete contexts, in particular in graph theory (see references \cite{hajnal,alon,kuhn}). Given a graph $G$ and a subgraph $H$ of $G$, an $H$-tiling in $G$ is a collection of vertex-disjoint copies of $H$ in $G$, that is, $G$ is tiled (covered) by disjoint copies of $H$, all the copies being \emph{identical} or \emph{isomorphic} in some relevant sense. The idea of tiling for finite fields is quite similar to the one in graph theory. The definition we adopt for the particular case of $\mathbb{F}_2^n$ considers the vectorial structure of the space (we consider translated copies of a given tile) but considering the Hamming cube as a graph, it coincides with the most usual definition of tiling of a graph (see reference \cite{balogh}).

\vspace{3pt}
\begin{defi}\cite{branko}\label{tiling1}
	A \textbf{tiling} of $\mathbb{F}_2^n $ is a pair $(D,C)$, where $D,C \subseteq \mathbb{F}_2^n$ are subsets satisfying 
	\[\displaystyle \bigcup_{\bc\in C} \bc+D =\mathbb{F}_2^n \ \ \mbox{and} \ \ (\bc+D) \cap(\bc'+D) =\emptyset,\]
	for all $\bc,\bc'\in C, \bc\not=\bc'$.

\end{defi}
 \vspace{3pt}
 
Despite the fact that the role of $D$ and $C$ are interchangeable, we shall call $D$ as a \textbf{tile} and $C$ as the \textbf{code}, since this is the role it will play in the context of coding theory. If $D$ is a polyhedromino, we say $(D,C)$ is a \textbf{poly-tiling} of $\mathbb{F}_2^n $. If $D$ is a convex polyhedromino, then $(D,C)$ is a \textbf{convex-tiling}.

 Notice that only translated copies of $D$ are considered, which is very reasonable in the context of TS-metrics, since in this case all the translated copies of the tile are isometric. Also, as we shall see, it is also reasonable the use of polyhedrominoes to tile $\Fn$. 
 
Since we are working with TS-metrics, a translation of a (convex) polyhedromino is also a  (convex) polyhedromino, so we may exchange a tiling $(D,C)$ by a  tiling $(D^\prime ,C^\prime )$ where $D^\prime = \bx +D,C^\prime =\bx +C$ and $\bzero \in D^\prime \cap C^\prime$. So, from here on, without loss of generality we assume that  $\bzero\in D$ and $\bzero\in C$.

It is trivial to see that given a tiling $(D,C)$, we have that $|D|\cdot |C|=|\mathbb{F}_2^n|$.
A trivial (and not interesting) way of obtaining a poly-tiling is to consider $I\subset [n]$ and letting $D_I=\{\bx =(x_1,\ldots ,x_n);x_i=0, i\in I\}$ and $C_I=\{\bx =(x_1,\ldots ,x_n);x_i=0, i\in [n]\setminus I\}$.

Our main reference is the work \cite[Cohen et al., 1995]{vardy} which adopts a different, but equivalent, definition of tiling. To present it, we set the notation $2D:=D+D=\{\bx +\by; \bx,\by\in D \}$.
\vspace{3pt}
 
\begin{defi}\cite{vardy}\label{tiling2}
The pair $(D,C)$ is a tiling of $\mathbb{F}_2^n$ if $D+C=\mathbb{F}_2^n$ and $2D\cap 2C=\{\bzero\}$, where both $D$ and $C$ contain the element ${\bzero}$.
\end{defi}

We now show that definitions \ref{tiling1} and \ref{tiling2} are equivalent. 

\vspace{3pt}

\begin{prop}
Definitions \ref{tiling1} and \ref{tiling2} are equivalent.
\end{prop}

\vspace{3pt}

\begin{IEEEproof}
Let $(D,C)$ be a tiling in the sense of Definition \ref{tiling1}. 
It means that $(c_i+D) \cap(c_j+D) =\emptyset$, for all $i\not= j$. Suppose there exists $0\not= \by\in 2D\cap 2C$. It means there are $\bx_1,\bx_2\in D$ and $\bc_1,\bc_2\in C$ such that $\bx_1+\bx_2=\bc_1+\bc_2$. It follows that $\bc_1-\bx_1=-\bc_2+\bx_2$ and since we are considering the binary case, we have that $\bc_1+\bx_1=\bc_2+\bx_2$. We note that $\bc_1+\bx_1=\bc_2+\bx_2\in (\bc_1+D)\cap (\bc_2+D)$. Since $\bc_1+\bc_2=\by\neq 0$ and the sum is binary, we get that $\bc_1\neq \bc_2$, a contradiction.

The reciprocal follows in the same manner and therefore, the two definitions are equivalent.
	\end{IEEEproof}

\subsection{Concatenation of tilings}\label{sec:concatenation}

Given $\ba=(a_1,a_2,\ldots,a_n)\in \mathbb{F}_2^n$ and $\bb=(b_1,b_2,\ldots,b_m)\in \mathbb{F}_2^m$, the \textbf{concatenation of the vectors}  $\ba$ and $\bb$ is $\ba\mid \bb:=(a_1,a_2,\ldots,a_n,b_1,b_2,\ldots,b_m)\in\Fnm$. Similarly, for $A\subset\Fn$ and $B\subset\Fm$, the set  $A\mid B=\{\ba\mid \bb; \ \ba\in A, \ \bb\in B\}\subset\Fnm$ is the \textbf{concatenation of the sets}
 $A$ and $B$.

\vspace{3pt}
The \textbf{rank of} a set $A\subset\Fn$ is the dimension of the vector subspace spanned by $A$, ie, $\rank (A)=dim\langle A \rangle$; the \textbf{rank of a tiling} $(D,C)$ is $\rank (D,C)=\rank (D)$. 

\vspace{3pt}

Given two tilings $(D_1,C_1)$ and $(D_2,C_2)$ of $\Fn$ and $\Fm$, respectively, it was proved in \cite{vardy} that the concatenation $(D_1\mid D_2,C_1\mid C_2)$ results in a tiling. If $\rank(D_1)=n$ and $\rank(D_2)=m$, then $\rank(D_1\mid D_2)=n+m$. 

\vspace{3pt}

As we shall see in Section \ref{sec:concatenation and perfect codes}, if $(D_1,C_1)$ and $(D_2,C_2)$ give rise to perfect codes on $\Fn$ and $\Fm$, the concatenated tiling also gives rise to a perfect code on $\Fnm$.

%Given tilings $(D_1,C_1)$ and $(D_2,C_2)$ of $\Fn$ and $\Fm$ respectively, it was proved in \cite{vardy} that:
%
%\begin{teo}[Theorem 7.5, \cite{vardy}] \label{teo:concat}
%Full-rank tilings of $\Fn$ exist for all sufficiently large $n$. 
%\end{teo}
%
%\vspace{3pt}
%
%\begin{IEEEproof}
%Let $(D_1,C_1)$ and $(D_2,C_2)$ be tilings of $\Fn$ and $\Fm$, respectively. Define $D=\{x_1\mid x_2; x_i\in D_i\}$, $C=\{c_1\mid c_2; c_i\in C_i\}$. 
%
%Then, clearly $2D\cap 2C=\{\bzero\}$ and $|D|.|C|=2^{n+m}$, which means that $(D,C)$ is a tiling of $\Fnm$. Now $\rank(D)=\rank(D_1)+\rank(D_2)$ and $\rank(C)=\rank(C_1)+\rank(C_2)$. Thus, if $(D_1,C_1)$ and $(D_2,C_2)$ are full-rank tilings then so is $(D,C)$. 
%\end{IEEEproof}
%
%\vspace{3pt}
%
%As proved in Theorem \ref{teo:concat}, the pair $(D,C)=(D_1\mid D_2,C_1\mid C_2  ) $ is a tiling of $\Fnm$, the \textbf{concatenated tiling}. 

\section{Obtaining perfect codes out of tilings}\label{sec:obtain}

Tilings are frequently studied in the context of graph theory and the Hamming cube $\Fn$ is a particular case of a graph. Tilings and perfect codes are two relevant research problems, that although distinct are  closely related. 
We start this part of the work by establishing this relation and then proceed as follows: 

\vspace{3pt}
In Section \ref{subsec:tiles}, we consider all the equivalence classes of tilings $(D,C)$ where $D=8$, which were characterized in \cite{vardy} and we determine each of those $D$ is a ball of a $\mathcal{TS}$-metric, giving rise to a perfect code $(D,C)$; in Section \ref{sec:large}, we present necessary and sufficient conditions for a tiling of maximal rank (as presented in \cite{vardy}) to be a TS-ball; in Section \ref{extensaotile} given a tiling $(D,C)$, with $D,C\subset\Fm$, we show how simple concatenations can lead tilings $(D^\ast,C^\ast)$ with $D^\ast,C^\ast\subset \Fnm$, $|D|=|D^\ast|$ and out of a $TS$-metric $d$ on $\Fm$ that turns $C$ into a perfect code we  define a metric $d^\ast$ on $\Fnm$ that does the same to $D^\ast$; finally, in Section  \ref{equivmetrics} we classify all TS-metrics that turn $D$ into a ball or equivalently, turn $C$ into a TS-perfect code.

\vspace{3pt}
The next proposition establishes a connection between tilings and perfect codes. 

Before we continue, we need some notations. We denote by $\be_i$ the (unique in a binary space) vector in $\Fn$ such that $\supp (\be_i)=\{i \} $ and call $\beta =\{ \be_1,\be_2,\ldots ,\be_n\}$ the \textbf{standard basis} of $\Fn$.

\vspace{3pt}
\begin{prop}\label{prop:poly}
	Given  a tiling $(D,C)$ of $\mathbb{F}_2^n$, suppose that $D=B_d(\bzero,r)$ for some  $d\in \mathcal{TS}(n)$. Then:
	
	\begin{enumerate}
		\item $D$ is a polyhedromino;
		\item $C$ is a $(d,r)$-perfect code. 
	\end{enumerate} 
\end{prop}

\begin{IEEEproof}
We remark that  the first part of the proposition demands the metric $d$ to respect support, while the second part demands it to be translation-invariant.
		\begin{enumerate}
			\item Since $D$ is a $d$-ball for some metric $d\in\mathcal{TS}$, we have that for every $\bx\in D$, if $\supp(\by)\subseteq \supp(\bx)$, then $\by\in D$. 
			We need to prove that there is a path $\gamma: \bx_0=\bx,\bx_1,\ldots \bx_{r-1},\bx_r=\by$ where $r=d_H(\bx,\by )$. 
		
			\vspace{3pt}	
			If $\supp (\bx)\subset \supp (\by)$ the problem is trivial, at every step we just adjoin a different vector of the basis $\beta$ which is contained in $\supp (\by)\setminus\supp (\bx):=\{t_1, t_2, \ldots ,t_r\}$, that is, we just define $\bx_i=\bx_{i-1}+\be_{t_i}$.% where $\{t_1, t_2, \ldots ,t_r\}=\supp (\by ) \setminus \supp (\bx)$.
		
			\vspace{3pt}
			If $\supp (\bx)\nsubseteq \supp (\by )$, then for every $t_i\in \supp (\bx) \setminus \supp (\by)$ we have that $d_H(\by , \bx+\be_{t_i})=d_H(\bx,\by)-1$ and $d_H(\bx , \bx+\be_{t_i})=1$. So, we choose $t_1\in \supp (\bx) \setminus \supp (\by)$ and set $\bx_1=\bx_0+\be_{t_1}$. 
		
		\vspace{3pt}
		
		We remark that in both cases we have that $\bx_1\in D$, since either $\supp(\bx_1)\subset\supp(\by)$ or $\supp(\bx_1)\subset\supp(\bx)$ with $\bx,\by\in D$ and we assumed that $D$ is a ball of a metric which respects support.
			\vspace{3pt}
			
			Now we proceed as before, considering $\bx_1$ instead of $\bx_0$. We set  $\bx_2=\bx_1+\be_{t_2}$, where $t_2\in \supp(\by)\setminus \supp(\bx)$ if $\supp (\bx_1)\subset\supp (\by)$ and $t_2\in \supp(\bx_1)\setminus\supp(\by)$ otherwise.
			\vspace{3pt}
			
			Since at each step we have that $d_H(\bx_i,\by )< d_H(\bx_{i+1},\by)$ we get a geodesic path from $\bx$ to $\by$ contained in $D$ and therefore, $D$ is a polyhedromino.  
			\item We have that $D$ is a ball and since $d$ is a translation-invariant metric, $c+D$ is also a ball. Since the pair $(D,C)$ is a tiling of $\Fn$ and $c+D$ is a ball for all $c\in C$, then $C$ is a $(d,r)$-perfect code. 
	\end{enumerate}
\end{IEEEproof}	

In the case where the conditions of the proposition holds, we say that the tiling $(D,C)$ \textbf{determines a TS-perfect code}.

\vspace{3pt}

\begin{exem}
		We  consider the trivial repetition code $C=\{\bzero,\mathbf{1}\}$. This is a $d_H$-perfect code (relatively to the Hamming metric) for every odd $n$. If $P=([n]\preceq )$ is any poset having a unique maximal element then $C$ is a $d_P$-perfect code (independently of $n$). If we assume that the maximal element is $i_0\in [n]$, then $B_{d_P}(\bzero ,n-1)=\{\bx\in\Fn; x_{i_0}=0  \}$ and $B_{d_P}(\mathbf{1},n-1)=\{\bx\in\Fn; x_{i_0}=1  \}$ are the two disjoint metric balls. It is important to remark that, as we have just seen, $C$ may be a $(d_H,r_H)$-perfect code and also to be a $(d_P,r_P)$- perfect code for a different poset metric. However, in this case, $r_H\neq r_P$. In our example of the trivial repetition code, for $n$ odd we have $r_H=\frac{n-1}{2}$ while $r_P=n-1$.
\end{exem}

%	\begin{prop}
%	Let $C$ be a $(d_H,1)$-perfect code. Then, $C$ can not be a $(d_P,r)$-perfect or $(d_{\FF},r)$-perfect code, using the same balls.   	
%\end{prop}
%
%\begin{IEEEproof}	
%	Let $B_H(\bzero,1)$ be the Hamming ball. Notice that, using the poset $Q=\{i\preceq j \ \mbox{and} \ j\preceq i\}$ we have that $d_P$ coincides with $d_H$. Then, this case will be excluded. Suppose that there exists a poset $P$ and $r\geq 1$ such that $B_P(\bzero,r)=B_H(\bzero,1)$. Since $P\not=Q$  then there is at least one $j$ such that $i\preceq j$. Since $\be_j\in B_H(\bzero,1)$ then $\be_j\in B_P(\bzero,r)$ and $\be_i+\be_j\in B_P(\bzero,r)$, but $d_H(\bzero, \be_i+\be_j)=2$. Therefore $B_P(\bzero,r)\not=B_H(\bzero,1)$ and $C$ is not a $(d_P,r)$-perfect code.      
%	
%	Now, suppose that $B_{\FF}(\bzero,r)=B_H(\bzero,1)$ for some covering $\FF$. Using the covering $\{\{1\},\{2\}, \{3\}, \ldots, \{n\} \}$, we have $d_{\FF}$ coincides with $d_H$, then this covering is not considered. Then, there exists at least a $A\subset \FF$ such that $|A|\geq 2$. With loss of generality, suppose that $A=\{i,j\}$, then $\be_i, \be_i$ and $\be_i+\be_j \in B_{\FF}(\bzero, r)$. But $d_H(\bzero, \be_i+\be_j)=2$ and then $e_i+e_j\notin B_H(\bzero,1)$. Therefore $B_{\FF}(\bzero, r) \not= B_H(\bzero,1)$.       	
%\end{IEEEproof}

\vspace{4pt}

The rest of this section is based on Proposition \ref{prop:poly} and in the work \cite{vardy}, where the authors classified tiles of $\mathbb{F}_2^n$ that are either ``small'' (and here small means $|D|\leq 8$) or of maximal rank. We remark that, since a tiling $(D,C)$ satisfies $|D||C|=2^n$,  by a ``small'' tile we mean a tile with cardinality $1,2,4$ or $8$. 

\vspace{3pt}

%
%In Sectionwe obtain all small tilings $(D,C)$ presented in \cite{vardy} and determines each of those $C$ is a TS-perfect code;  In Section we give necessary and sufficient conditions for a tiling of large rank presented in \cite{vardy} to determine a TS-perfect codes; In Section, given a perfect code $(D,C)$ with respect to a metric $d$ and with $D\subset\mathbb{F}_2^s$, we present a systematic way to extend $d$ into a metric $d^\ast$ on $\Fn$ which turns the extension of $(D,C)$ to $\Fn$ to be a perfect code; Finally, in Section we classify all TS-metrics that turn $D$ into a ball or equivalently, turn C into a TS-perfect code.    

%   we characterize the properties of a ball in a TS-metric. On the second part, we list all the small tilings described in \cite{vardy} and determine which one is a ball (or not) for some TS-metric. In \cite{vardy}, they characterize all tiles of $\mathbb{F}_2^n$ up to size 8. In the case where the tiling cardinality is equal to 8 elements, they separate the tiles by rank, which variate from 3 to 7. \\

\subsection{Classifying small tiles that determine TS-perfect codes}\label{subsec:tiles}

We start this point by giving a single example of many ways how to turn a tile into a metric ball.

\begin{exem}
Let $D=\{\bzero, \be_1, \be_2,,\be_1+\be_2\}$ be a tile of $\mathbb{F}_2^3$. Then, considering the posets, $P_1,P_2$ represented, respectively, by the Hasse diagrams

\begin{figure}[htb]
\begin{center}
\includegraphics[width=1.5cm]{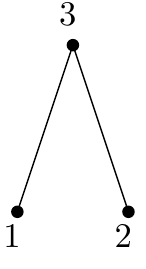}\hspace{2.5cm}
				\includegraphics[width=1.5cm]{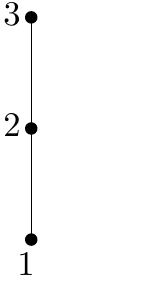}
\end{center}
		\label{fig:dir}
\end{figure}
we have that $D=B_{d_{P_1}}(\bzero,3)=B_{d_{P_2}}(\bzero,3)$. 

%Consider the coverings $\FF_1=\{1,2,3\}$, $\FF_2=\{\{3\},\{1,2\}\}$ and $\FF_3=\{\{1\},\{2\},\{3\}\}$. Thus, $D=B_{d_{\FF_i}}(\bzero,i)$.  
\end{exem}

\vspace{3pt}

We start considering the possibilities for ``very small'' tiles, that is, tiles with $2$ or $4$ elements.

\begin{table*}[b]
	\centering
	\begin{tabular}{|c|c|c|}
		\hline
		$Rank$ & Elements & Counter-example\\
		\hline
		$4$ & ${\bzero}, \be_1, \be_2, \be_3, \be_4, \be_1+\be_2,\be_1+\be_3,\be_1+\be_2+\be_3$ & $\be_2+\be_3$ \\ \hline
		$4$ & ${\bzero}, \be_1, \be_2, \be_3, \be_4, \be_1+\be_2, \be_1+\be_2+\be_3, \be_1+\be_2+\be_4$ & $\be_1+\be_3$  \\ \hline
		$4$ & ${\bzero}, \be_1, \be_2, \be_3, \be_4, \be_1+\be_2+\be_3, \be_1+\be_2+\be_4,\be_1+\be_3+\be_4$ & $\be_1+\be_2$\\ \hline
		$4$ & ${\bzero}, \be_1, \be_2, \be_3, \be_4, \be_1+\be_3, \be_1+\be_2+\be_3,\be_1+\be_2+\be_4$ & $\be_2+\be_4$\\ \hline
		$5$ & ${\bzero}, \be_1, \be_2, \be_3, \be_4, \be_5, \be_1+\be_2+\be_3, \be_1+\be_2$ & $\be_1+\be_3$ \\\hline 
		$5$ & ${\bzero}, \be_1, \be_2, \be_3, \be_4, \be_5, \be_1+\be_2+\be_3, \be_4+\be_5$ & $\be_1+\be_2$\\\hline 
		$5$ & ${\bzero}, \be_1, \be_2, \be_3, \be_4, \be_5, \be_1+\be_2+\be_3, \be_1+\be_2+\be_4$ & $\be_1+\be_2$\\\hline
		$6$ & ${\bzero}, \be_1, \be_2, \be_3, \be_4, \be_5, \be_6, \be_1+\be_2+\be_3$ & $\be_1+\be_2$ \\\hline
		$6$ & ${\bzero}, \be_1, \be_2, \be_3, \be_4, \be_5, \be_6, \be_1+\be_2+\be_3+\be_4+\be_5+\be_6$ & $\be_1+\be_2$\\\hline	
	\end{tabular}
	\caption{Tiles of type 1}
	\label{tiles}
\end{table*}

\vspace{3pt}

\begin{prop}\label{prop1}
Let $B=B_d({\bzero},r)\subseteq\mathbb{F}_{2}^{n}$ be a TS-ball with $2$ or $4$ elements. Then, $B$ is one of the following:
\begin{align*}
	B_1 & = \left\{ \bzero,\be_{i}\right\},\\
	B_{2}  & = \left\{  \bzero,\be_{i},\be_{j},\be_{k}\right\}, i,j,k \ \text{distincts},\\
	B_{3}  & =\left\{  \bzero,\be_{i},\be_{j},\be_{i}+\be_{j}\right\}, i\neq j.
	\end{align*}
\end{prop}

\begin{IEEEproof}
	All convex polyhedrominoes of  size $2$ or $4$ are the listed ones: $B_1, B_2$ and $B_3$. Hence, by Proposition \ref{prop:poly} these are all the possible candidates for a TS-ball with $2$ or $4$ elements. We need to show that these are indeed balls of some TS-metric. In fact, each one may be realized as a ball of  a poset metric, determined, respectively, by the \underline{non-trivial} sets of relations: 
	\begin{align*}
	B_1: \hspace{6pt} & \{i\preceq l;\forall l \neq i  \},    \\
	B_2: \hspace{6pt} & \{ t\preceq l; \forall t\in \{i,j,k\}, l\in [n]\setminus \{i,j,k\}   \},     \\
	B_3: \hspace{6pt} &   \{ t\preceq l; \forall t\in \{i,j\}, l\in [n]\setminus \{i,j\}   \}.         
	\end{align*}
	
		\end{IEEEproof}

\vspace{3pt}
 
Notice that, if the condition of respecting support was not required, there would exist other polyhedrominoes with 2 or 4 elements. For example, $D=\{\bzero, \be_i, \be_{i}+\be_j, \be_i+\be_j+\be_k\}$ is a polyhedromino, but it is not a ball, since $D$ does not respect support.

\vspace{3pt}

In \cite{vardy} there is a complete classification of small tiles of $\mathbb{F}_2^n$. In that work, the authors consider tilings $(D,C)$ of $\Fn$ assuming that $|D|=8$ and that $D$ has full rank, in the sense that the linear space spanned by $D$ has dimension $n$. Considering all the possibilities presented in this classification, there is a total of $193$ different tiles. However, many of those tiles are equivalent, in the sense that they can be obtained one from the other by a permutation of the coordinates. To be more precise, two tiles $D_1,D_2\subset \Fn$ are said to be \textbf{equivalent} if there is a permutation $\sigma\in S_n$ such that $\sigma (D_1)=D_2$, where the action is on the coordinates of each element of the tile: for $\bx=(x_1,x_2,\ldots ,x_n)$ we have $\sigma(\bx)=(x_{\sigma(1)} ,x_{\sigma(2)},\ldots ,x_{\sigma(n)}  )$. By carefully looking at each case and finding an appropriate permutation, it is possible to reduce the list to $15$ equivalence classes, which are presented in the next Proposition. The proof is lengthly (it considers $193$ cases) but simple, so it is omitted in this work. The details, that is, the list of tiles and permutations, can be found in \cite{akemi}, which should be considered as an appendix of this work.

\vspace{3pt}

\begin{prop}
Every tile $D$ presented in \cite{vardy}, with $|D|=8$ is equivalent to one of the tiles in Tables \ref{tiles}, \ref{tilesball} and \ref{tilesball2}. 
\end{prop}

\begin{table*}[b]
	\centering
	\begin{tabular}{|c|c|c|c|c|}
		\hline
		Tile & Rank & Elements & Radius & Non trivial relations of the Poset \\
		\hline
		$D_1^3$ & $3$ & ${\bzero}, \be_1, \be_2, \be_3, \be_1+\be_2, \be_1+\be_3,\be_2+\be_3, \be_1+\be_2+\be_3 $ & 3 & $P_1:1\preceq 2\preceq 3$, \\\hline
		$D_1^7$ & $7$ & $\bzero, \be_1, \be_2, \be_3, \be_4, \be_5, \be_6, \be_7$ & 1 & $P_2:$ only trivial relations\\ \hline
	\end{tabular}
	\caption{Tiles of type 2}
	\label{tilesball}
\end{table*}

\begin{table*}[b]
	\centering
	\begin{tabular}{|c|c|c|c|c|}
		\hline
		Tile & Rank & Elements & Radius & Combinatorial metric\\
		\hline
		$D_1^4$ & 4 & ${\bzero}, \be_1, \be_2, \be_3, \be_4, \be_1+\be_2, \be_1+\be_3,\be_1+\be_4$ & 1 & $\mathcal{F}_1=\{\{1,2\},\{1,3\},\{1,4\}\}$\\ \hline
		$D_2^4$ & 4 & ${\bzero}, \be_1, \be_2, \be_3, \be_4, \be_1+\be_2, \be_1+\be_3,\be_2+\be_3$ & 1 & $\mathcal{F}_2=\{\{1,2\},\{1,3\},\{2,3\},\{4\}\}$ \\ \hline
		$D_1^5$ & 5 & ${\bzero}, \be_1, \be_2, \be_3, \be_4, \be_5, \be_1+\be_2, \be_1+\be_3$ & 1 & $\mathcal{F}_3=\{\{1,2\},\{1,3\},\{4\},\{5\}\}$\\\hline
		$D_1^6$ & 6 & ${\bzero}, \be_1, \be_2, \be_3, \be_4, \be_5, \be_6, \be_1+\be_2$ & 1 & $\mathcal{F}_4=\{\{1,2\}, \{3\}, \{4\}, \{5\}, \{6\}\}$ \\\hline
	\end{tabular}
	\caption{Tiles of type 2}
	\label{tilesball2}
\end{table*}

If we wish to determine which of the 193 tiles presented in \cite{vardy} give rise to a TS-perfect code, Proposition \ref{equiv} ensures that it is enough to check it for the 15 cases presented in the tables  \ref{tiles}, \ref{tilesball} and \ref{tilesball2}. % ones, it is necessary to state that if $D$ is a ball then the equivalent tile will be a ball, and the converse is also true. This fact is stated in the next proposition.  

\vspace{3pt}

\begin{prop}\label{equiv}
Let $D$ be a tile and $\sigma\in S_n$ a permutation of $n$ and let $\sigma(\bx)=(\bx_{\sigma(1)},\ldots,\bx_{\sigma(n)})$. Then $D=B_d(\bzero,r)$ \emph{iff} $\sigma(D)=B_{d_\sigma}(\bzero,r)$, where $d_\sigma$ is the metric determined by the weight $\omega_\sigma$,  defined by $\omega_{\sigma}(\bx_{\sigma(1)},\ldots,\bx_{\sigma(n)}):=\omega(\bx_1,\ldots,\bx_n)$.
\end{prop}

The proof is needles, since $d_\sigma$ is, by definition, the metric induced by the map $\sigma$, a standard procedure.

\vspace{3pt}

The list of tiles representing the equivalence classes is divided into three separated tables because they play different roles.

Table \ref{tiles} contains all tiles that cannot be a ball for any TS-metric. We remark that, if $D=B_d(\bzero,r)$ is a ball for some TS-metric and   $\bx\in B_d(\bzero,r)$, then $\by\in B_d(\bzero,r)$ for all  $\by\in\Fn$ such that $\supp(\by)\subseteq \supp(\bx)$. This simple remark makes possible to eliminate all the $9$ tiles of Table \ref{tiles}, since they do not satisfy this condition.  As an example, consider the tile $$D=\{{\bzero}, \be_1, \be_2, \be_3, \be_4, \be_1+\be_2, \be_1+\be_3,\be_1+\be_2+\be_3\},$$ in the first line of the table. Note that  $\supp(\be_2+\be_3)\subset \supp(\be_1+\be_2+\be_3)$ but $\be_2+\be_3\notin D$, what leads to a contradiction: If $D$ was to be a metric ball, we would  have $\omega(\be_2+\be_3) \leq \omega(\be_1+\be_2+\be_3)$. 

On each row of the table, in the last column we present a vector which will lead to a similar contradiction. 

\vspace{3pt}
The remaining $6$ tiles are presented in Tables \ref{tilesball} and \ref{tilesball2}. They are denoted by $D_{j}^s$, where $s=\rank(D_j^s)$ and $j$ is just a counting index. These six tilings lead to perfect codes since the basic tiles can be realized as a metric ball of a TS-metric: The tiles on Table \ref{tilesball} are balls of some poset metric while the tiles in Table \ref{tilesball2} are realized by combinatorial metrics. The proof of this fact is constructive, we just present (in the last column of each table) a poset structure (actually its non-trivial relations) or an appropriate covering. 

\vspace{3pt}
As a consequence, we have the following:

\vspace{3pt}

\begin{teo}\label{nottile}
	A tile $D\subset \Fn$ of rank $n$ and cardinality $8$ is a metric ball of some TS-metric \emph{iff} it is equivalent to a tile presented in  Table \ref{tilesball} or \ref{tilesball2}. 
\end{teo}

%\begin{IEEEproof}\textcolor{blue}{
%The proof consists in notice that if $D$ is not equivalent to the ones in Table \ref{tilesball} or \ref{tilesball2}, then $D$ is equivalent to a tile in Table \ref{tiles}, and in the third column of table, we find the counter-examples showing that $D$ can not be a TS-ball. }	
%\end{IEEEproof}
%
%\vspace{3pt}

%
%To show that the remaining tiles give rise to a TS-perfect code, we need to find a TS-metric which turns them into a metric ball. The proof of the next theorem, Theorem \ref{teo:TS}, is actually the last column of the tables, where we present a poset metric (Table \ref{tilesball}) or a combinatorial metric (Table \ref{tilesball2}) that turns the tile into a ball. 

Due to Proposition \ref{prop:poly}, Theorem \ref{nottile} concerning tilings can be re-stated as a result about the existence of perfect codes.

\vspace{3pt}

\begin{teo}\label{teo:TS}
Given a tiling $(D,C)$ of $\Fn$, where $D$ has rank $n$ and cardinality $8$, the code $C$ is a TS-perfect code \emph{iff} $D$ is equivalent to a tile presented in  Table \ref{tilesball} or Table \ref{tilesball2}.
\end{teo}

\begin{IEEEproof}
It follows straightforward from Theorem \ref{nottile} and Proposition \ref{prop:poly}.
\end{IEEEproof}

\vspace{3pt}

\begin{rema}
	The tiles $D$ listed in Tables \ref{tilesball} and \ref{tilesball2} are considered as subsets of $\mathbb{F}_2^s$, where $s=\rank(D)$. In Section \ref{extensaotile} we show a process used to extend them to $\Fn$, $n\geq s$.
\end{rema}

\subsection{Classifying tiles with large rank that determine TS-perfect codes}\label{sec:large}\label{sub:large}

In the previous section, we presented small tilings of the binary space. Despite the fact that the tiles had full rank, the rank was always small, since $\rank(D)\leq |D|-1$. Now, we give necessary and sufficient to a tile of  rank $n$ and cardinality $n+2$ to determine a TS-perfect code. For that, we use a proposition proved in \cite[Proposition 4.5]{vardy} which states that a set $D_n(\bx)= \{\be_i;i\in [n]  \} \cup \{ \bzero, \bx\}  $, for some $\bx\in\Fn$ with $\omega_H(\bx)\geq 2$ is a tile \textit{iff} $\omega_H(\bx)\notin \{n-1,n-2 \}$. We shall determine a necessary and sufficient condition for it to define a TS-perfect code.
\vspace{3pt}

\begin{prop}
Suppose that $(D_n(\bx),C_n(\bx))$ is a tiling of $\Fn$. Then, there is a TS-metric that turns it into a perfect code \textit{iff} $\omega_H(\bx)=2$. 
\end{prop}

\begin{IEEEproof}
	If $\omega_H(\bx)>2$, then $D_n(\bx)$ cannot be a ball in a metric that respects support, since in this case there would be some subset $A\subset\supp (\bx)$ with $1<|A|<\omega_H(\bx)$ and the vector $\bx_A$ defined by $\supp (\bx_A)=A$ is not contained in $D_n(\bx)$. 
	
	For $\omega_H(\bx)=2$, we have that $\bx=\be_j+\be_k$, for some $j\neq k$. We define $\mathcal{F} = \{  \{i\} ; i\in [n] \}  \cup \{ j,k \}$ and we have that $D_n(\bx)=B_{d_\mathcal{F}}(\bzero,1)$ and, by Proposition \ref{prop:poly} it follows that $(D_n(\bx),C_n(\bx))$ is a $d_\mathcal{F}$-perfect code.
		\end{IEEEproof}

\subsection{Extending tilings from $\mathbb{F}_2^s$ to $\Fn$}\label{extensaotile}

\vspace{3pt}

In Sections \ref{subsec:tiles} and \ref{sub:large}, we considered tilings $(D,C)$ of $\mathbb{F}_2^s$ where $s=\rank (D)$. Since $\mathbb{F}_2^s$ can be seen as a linear subspace of $\Fn$ for $n\geq s$, we can extend that to a tiling $(D^\ast ,C^\ast)$ of $\Fn$. We denote $\bzero_l$ the null element in $\mathbb{F}_2^l$ and let $D^\ast =D\mid \bzero_{n-s}$ and $C^\ast=C \mid \mathbb{F}_2^{n-s}$. As can be found in  \cite{vardy} we have that $(D^\ast ,C^\ast)$ is a tiling of $\Fn$. We remark that since we are concatenating $D$ with the null space, the cardinality   and rank of $D^\ast$ are the same as those of $D$ ($|D^\ast |=|D|$ and $\rank(D^\ast)=\rank(D)$).

As a code construction, this is a rather not interesting situation. Nevertheless, in Section \ref{sec:concatenation and perfect codes} we shall present some non-trivial concatenations. For this reason,  we shall see here that a TS-metric $d$ which turns $(D,C)$ into a perfect code can be extended to do the same for the concatenated code, that is,  we can extend it to a TS-metric $d^\ast$  which turns $D^\ast$ into a metric ball $B_{d^\ast}(\bzero,r^\prime)$ in $\Fn$. 

\vspace{3pt}

\begin{teo}\label{teo:exttiles}
	Given $D=B_{d}(\bzero,r)$, $d\in \mathcal{TS}(s)$, there is  $d^\ast\in \mathcal{TS}(n)$ such that $D^{\ast}=B_{d^\ast}(\bzero,r)$.
\end{teo}

\begin{IEEEproof}
Given a  weight $\omega$ on $\mathbb{F}_2^s$, let $M(\omega) = \max\{\omega(\bx);\bx\in D  \} $. We define, for $\bx\in\Fn$, $n\geq s$
\[
\omega_{n,s} (\bx)= \begin{cases} \omega (\bx) \text{ if } \supp(\bx)\subset [s]\\
M(\omega )+1 \text{ otherwise}.\end{cases}
\]

It is not difficult to see that $\omega_{{n,s}} (\bx)$ is a weight. Let $d$ and $d_{s,n}$ be the metrics determined by $\omega$ and $\omega_{{n,s}}$ respectively. It is not difficult to prove that  $d$ respects the support of vectors \textit{iff}  $d_{s,n}$ does it. Moreover,
	\[
B_{d_{(n,s)}}(\bzero,r) =	B_d(\bzero,r)\mid \{ \bzero_{n-s} \}
	\]	
for every $r\leq M(\omega )$. In other words, if $(D,C)$ determines a TS-perfect code, so does $(D^\ast ,C^\ast)$.
\end{IEEEproof}

\vspace{3pt}
\begin{rema}
In the two cases considered in Table \ref{tilesball}, where the metrics were determined by a poset $P$ over $[s]$, it is possible to   extend it to a metric defined by a poset $P^\ast$ over $[n]$, leading to a more natural construction, as follows:
	 $P_1^{\ast}$ is defined by the (non-trivial) relations  $1\preceq 2\preceq 3$ and $3\preceq i$ for all $i\geq 4$. The poset  $P_2^{\ast}$ is defined by the (non-trivial) relations $i\preceq j$ for all $ i\leq 7 < j$. These are actually the minimal poset metrics which extend the original ones and it is not difficult to classify all the poset extensions that do it.
	 
	 %For the cases in Table \ref{tilesball2}, the extension follows by directly applying Theorem \ref{teo:exttiles}.
\end{rema}

\bigskip

\subsection{Classifying the TS-metrics which turn a tiling into a perfect code}\label{equivmetrics}

If $(D,C)$ determines a perfect code, by definition there is $d\in\mathcal{TS}$ that turns $D$ into a metric ball. Actually, there are infinitely many such metrics (takes, for example, any positive multiple of $d$), so when we wish to classify all such metrics, we mean up to an adequate equivalence relation. The most natural equivalence relation in the context of coding theory is to say that two metrics on $\Fn$ are equivalent if they determine the same minimum distance decoding for every code $C\subset \Fn$ and every received message $\bx\in\Fn$. To be more precise:

\begin{defi}\label{def:equiv}
Two metrics (or distances) $d_1$ and $d_2$ defined over $\mathbb{F}_2^n$ are \textit{decoding equivalent}, denoted by $d_1\sim d_2$, if 
\[\argmin\{d_1(\bx,\bc): \bc\in C\} = \argmin\{d_2(\bx,\bc): \bc\in C\},\]
for any code $C\subseteq \mathbb{F}_2^n$ and any $\bx\in \mathbb{F}_2^n$.
\end{defi}
\vspace{3pt}

It is not difficult to see that $d_1\sim d_2$ \textit{iff} $d_1 (\bx,\by)<d_1 (\bx,\bz)\iff d_2 (\bx,\by)<d_2 (\bx,\bz)$, for all $\bx,\by,\bz\in\mathbb{F}_2^n$. 
Details about this equivalence relation can be found in \cite{rafael} and \cite{rafael2}. 

%In \cite{rafael}, the authors presented a manner to verify which are the equivalent metrics to a given one. In this section, we use their result to show which are the equivalent metrics to the metrics that made the tiles (in Tables \ref{tilesball} and \ref{tilesball2}) a ball.

Let $M\subset \mathbb{F}_2^{N}\times \mathbb{F}_2^{N}$, $N=2^n$ be the \textbf{distance matrix} of a metric $d\in\mathcal{TS}(n)$, defined by  $m_{\bx,\by}:=d(\bx,\by)$. Our goal is to determine necessary and sufficient conditions (on the matrix $M$) to determine a TS-metric that turns a tiling $(D,C)$ into a perfect code. This is what is done in the next theorem.

%Let $D$ be a tile that is determined by a TS-metric (those in Tables \ref{tilesball} and \ref{tilesball2} and let  $d$ be the metric determined in Proposition \ref{exttiles} which turns $D$ into a metric ball of radius $r$ in $\mathbb{F}_2^s$. Consider a matrix  
\vspace{3pt}

\begin{teo}
		Let $(D,C)$ be a tiling of $\Fn$. Let $d$ be a TS-metric for which $D=B_d(\bzero,r)$.	Let
		 $M =(m_{\bx,\by})\subset \mathbb{F}_2^{N}\times \mathbb{F}_2^{N}$ be a $N\times N$ matrix, with $N=2^n$, satisfying the following conditions:
		\begin{enumerate}
			\item[C1)]\label{c1} $m_{\bx,\bzero} = d(\bx,\bzero)$ for $\bx\in D$.
			\item[C2)] $m_{\bx,\bzero}>r$ for $\bx\notin D$.
			\item[C3)] $m_{\bx,\by}=m_{\by-\bx,\bzero}$ for all $\bx,\by\in\mathbb{F}_2^N$.
			
		\end{enumerate}
	
	Then, the following holds:
\begin{enumerate}
	\item[i)] The matrix $M$ defines a distance $dist_M$ which is decoding equivalent to a metric $d_M$  that is a translation-invariant metric. 
	\item[ii)] The tile $D$ is a metric ball of the metric $d_M$, to be more precise, $D=B_{d_M}(\bzero,r)$.
	\item[iii)] It is possible to choose the values of $m_{\bx,\by}>r$ for $\bx\notin D$ in such a way that the metric $d_M\in \mathcal{TS}(N)$.
	\item[iv)] Any TS-metric $d'$ which turns $D$ into a metric ball is equivalent to a metric described by a matrix $M$ satisfying conditions C1, C2, C3.

\end{enumerate}

\end{teo}

\begin{IEEEproof}  
	\begin{enumerate}
		\item[i)] Since we are considering the binary space, we have that $\bx-\by=\by-\bx$, then $M$ is symmetric. Moreover, conditions C1 and C2 ensures the positivity condition on the first row of the matrix. Condition C3 ensures the positivity on the other rows of $M$. So, we have that $M$ determines a distance. But on a finite space, any distance is equivalent to a metric (see \cite[Chapter 1.1]{deza2009}) and we have that $M$ defines a distance $dist_M$ which is decoding-equivalent to a metric $d_M$. The translation invariance follows from the fact that the first row determines all the others (Condition C3).
		\item [ii)] For all $\bx \in D$, we have $m_{\bx, \bzero}=d(\bx,\bzero)\leq r$ since $D=B_{d}(\bzero,r)$. Then, condition C1 ensures that  $d_M(\bx,\bzero)\leq r$ for all $\bx \in D$ and condition C2 ensures $d_M(\bx,\bzero)> r$ for all $\bx \notin D$, therefore $D=B_{d_M}(\bzero,r)$.
		\item [iii)] This item is made constructively. For all $x\notin D$, if $\supp(\bx)\subseteq \supp(\bz)$ then take $m_{\bzero,\bx}\leq m_{\bzero,\bz}$. Then, $d_M$ respects support. On item i), it was proved that $d_M$ is a translation-invariant metric, thus $d_M\in \mathcal{TS}(N)$.
		\item [iv)] It follows from the algorithm presented in \cite{rafael} to obtained a reduced form of a metric that any two metrics satisfying those conditions have the same reduced form and hence are equivalent.
	\end{enumerate}\end{IEEEproof}

\section{Concatenation of tilings: extending perfect codes to larger dimensions}\label{sec:concatenation and perfect codes}

In this section, we present some constructions to obtain new perfect codes out of a given pair of perfect codes. The main tool to achieve this goal is the concatenation of tiles. Notice that the extension made in Section \ref{extensaotile} is a particular (and trivial, concatenation  with the null space) case of what will be presented in this section.

Since we are working with poly-tilings, the first step is to prove that the concatenation of poly-tilings results in a poly-tiling. That is what is stated in the next two results.
\vspace{3pt}

\begin{prop}\label{polyhe}
Let $D_1\subseteq \mathbb{F}_2^n$, $D_2\subseteq\Fm$ and let  $D=D_1\mid D_2\subset \mathbb{F}_2^{n+m}$ be the concatenation of $D_1$ and $D_2$. Then, $D$ is a polyhedromino \textit{iff} $D_1$ and $D_2$ are polyhedrominoes.
\end{prop}

\begin{IEEEproof}
	Suppose that $D_1$ and $D_2$ are polyhedrominoes. Let $\bx,\by\in D$, $\bx=\bx_1\mid \bx_2$ and $\by=\by_1\mid \by_2$. We need to prove that there exists a geodesic path connecting $\bx$ to $\by$. Since $D_1$ is a polyhedromino there exists a geodesic path $\gamma_1\subset D_1$ connecting $\bx_1$ to $\by_1$. So, we can use $\gamma_1$ to connect $\bx_1\mid \bx_2$ to $\by_1\mid \bx_2$ in the following way: define $\gamma_1'=\{\bw\mid \bx_2; \bw\in \gamma_1\}$. We remark that a path is not only a set of points in the Hamming cube, but an \emph{ordered} set of points. Using this set notation for $\gamma_1'$ we are actually considering on it the order determined by $\gamma_1$. 
	Recalling that a geodesic path is characterized by the fact that its length equals the Hamming distance between any pair of its points,  since $\gamma_1$ is a geodesic path then so is $\gamma_1'$. Hence, we have that $\gamma_1'$ is a geodesic path connecting $\bx_1\mid \bx_2$ to $\by_1\mid \bx_2$. Similarly, since $D_2$ is a polyhedromino, there exists $\gamma_2\subset D_2$ connecting $\bx_2$ to $\by_2$. Define $\gamma_2'=\{\by_1\mid \bz; \bz\in \gamma_2\}$, then $\gamma_2'$ connects $\bx_1\mid \bx_2$ to $\by_1\mid \by_2$. Defining $\gamma:=\gamma_1'\cup\gamma_2'=\{\bw\mid \bx_2; \bw\in \gamma_1\}\cup \{\by_1\mid \bz; \bz\in \gamma_2\}$ we have that  $\gamma$ is a geodesic path connecting $\bx$ to $\by$ and thus $D$ is a polyhedromino.

	Suppose now that $D$ is a polyhedromino. By definition, we have $D=\{\bx_1\mid \bx_2; \bx_1\in D_1, \bx_2\in D_2\}$. Consider $\bx_1, \by_1 \in D_1$, then there exists $\bx_2\in D_2$ such that $(\bx_1,\bx_2), (\by_1,\bx_2)\in D$. Since $D$ is a polyhedromino, there exists $\gamma \subset D$, $\gamma=\{\bv\mid \bx_2; \bv\in D_1\}$ connecting $(\bx_1,\bx_2)$ to $(\by_1,\bx_2)$. Define $\gamma'=\{\bv; \bv\mid \bx_2\in\gamma\}\subset D_1$ and use it to connect $\bx_1$ to $\by_1$. Then, we have $D_1$ is a polyhedromino. In the same way we can prove that $D_2$ is a polyhedromino. 
\end{IEEEproof}

\vspace{3pt}

\begin{rema}
	If $D_1$ and $D_2$ are convex polyhedrominoes, the concatenation $D$ is not, necessarily, a convex polyhedromino. For example, let $D_1=D_2=\{000,100,010,001\}$ be convex polyhedrominoes in $\mathbb{F}_2^3$. Consider $\bx=100100,\by=010010\in D_1\mid D_2\subset \mathbb{F}_2^6$. Notice that $d_H(\bx,\by)=4$ hence every path $\gamma\subset \mathbb{F}_2^6$ connecting $\bx$ to $\by$ with $|\gamma|=4$ is a geodesic path. Consider $\gamma_1$ defined by $\gamma_1: 100100 \rightarrow 100000 \rightarrow 110000 \rightarrow 110010 \rightarrow 010010$. We have $|\gamma|=4$ (then $\gamma$ is a geodesic path), but $110000,110010 \notin D$ and thus $\gamma\not\subset D$. Therefore, $D$ is not a convex polyhedromino. However, the converse is true and it is stated in the next proposition.
\end{rema}

\begin{prop}
	Let $D=D_1\mid D_2\subset \mathbb{F}_2^{n+m}$. If $D$ is a convex polyhedromino then $D_1$ and $D_2$ are convex polyhedrominoes.
\end{prop}

\begin{IEEEproof}
	 Proposition \ref{polyhe} ensures that both $D_1$ and $D_2$ are polyhedrominoes, we just need to prove the convexity. 
	 We let $\bx,\by\in D_1$ and let $\gamma$ be a geodesic path connecting $\bx$ to $\by$. We need to prove that $\gamma\subset D_1$. We let $\bz\in D_2$ and define $\gamma':=\{ \gamma\mid \bz \}$. It is immediate to see that $\gamma'$ is a geodesic path connecting $\bx\mid\bz$ to $\by\mid \bz$. Since we are assuming that $D=D_1\mid D_2$ is convex, we have that $\gamma'\subset D$ and it follows that $\gamma\subset D_1$. The case for $D_2$ follows in the same manner, by considering a path $\gamma'=\bz\mid\gamma$ with $\bz\in D_1$ and $\gamma\subset D_2$.
%Therefore, $D_1$ and $D_2$ are convex polyhedrominoes. 
\end{IEEEproof}

\vspace{3pt}

In \cite[Theorem 7.5]{vardy}, it was shown that given two tilings $(D_1,C_1)$ and $(D_2,C_2)$ of $\mathbb{F}_2^n$ and $\mathbb{F}_2^m$, respectively, the concatenation $(D,C)=(D_1\mid D_2,C_1\mid C_2)$ is a tiling of $\mathbb{F}_2^{n+m}$. The same holds for poly-tilings. From this and Proposition \ref{polyhe}, we have the following:

\vspace{3pt}
\begin{coro}\label{concatenation}
	Let $(D_1,C_1)$ and $(D_2,C_2)$ be poly-tilings of $\mathbb{F}_2^n$ and $\mathbb{F}_2^m$, respectively. Then, $(D_1\mid D_2,C_1\mid C_2)   $ is a poly-tiling of $\Fnm$ \textit{iff} $(D_1,C_1)$ and $(D_2,C_2)$ are poly-tilings.

\end{coro}

\subsection{Extension of TS-perfect codes}

We proved (Corollary \ref{concatenation}) that the concatenation of two poly-tilings results in a poly-tiling. But, what happens in the case of \emph{convex} poly-tilings, which give rise to perfect codes? If two tilings determine  perfect codes, then the concatenation will be a perfect code? The answer is affirmative and we present two different constructive results. The first one, in  Theorem \ref{teo:tilingconca}, is in a more restrictive setting, where we consider the concatenation of tiles that are balls of the \emph{same} radius of two arbitrary TS-metrics.  In  Theorem \ref{teo:tilingposetconcat} we may consider balls of \emph{different} radii. Some results are valid only for combinatorial metrics and they will be show in the next section.  

Notice that the concatenation of two sets can be seen as a direct  product between them. Then, it would be natural to consider the product metric. But, in a general case, the concatenated tile $D$ is not  a metric ball in the product metric. For that reason, we define other metrics to accomplish our goal.   

From here on,  given $\bx\in\Fnm$, express $\bx:=\bx_1\mid \bx_2$, where $\bx_1\in\Fn$, $\bx_2\in\Fm$.

\vspace{3pt}

\begin{lema}\label{lema}
 Consider two metrics $d_1, d_2$ defined on $\Fn$ and $\Fm$, respectively, and define  $d_{\max}(\bx,\by):=max\{d_1(\bx_1,\by_1),d_2(\bx_2,\by_2)\}$. 
Then $d_{\max}$ is a metric on  $ \Fnm$ and $d_1\in\mathcal{TS}(n)$, $d_2\in\mathcal{TS}(m)$  implies~$d_{\max}\in\mathcal{TS}(m+n)$.
\end{lema}

\begin{IEEEproof}
The proof follows directly from the definition of a metric. The only sensitive points to pay attention are the following: 1) A translation on $\Fnm$ by a vector $\bx$ can be seen as the composition of the translation by the vector $\bx_1\mid \bzero_m$ followed by the translation by $\bzero_n\mid\bx_2$, where $\bx=\bx_1\mid\bx_2$ and $ \bzero_m\in\Fm,\bzero_n\in\Fn$; 2) $\omega_H(\bx)=\omega_H(\bx_1)+\omega_H(\bx_2)$.
\end{IEEEproof}

\vspace{3pt}
Now we consider the concatenation of two balls with same radius. 

\begin{teo}\label{teo:tilingconca}
	Let $(D_1,C_1), (D_2,C_2)$ be poly-tilings of $\mathbb{F}_2^n$ and $\mathbb{F}_2^m$, respectively. Suppose that $D_1=B_{d_{1}}({\bzero},r)$ and $D_2=B_{d_{2}}({\bzero},r)$, where $d_{1}, d_{2}$ are TS-metrics. 
		Let $(D,C) =(D_1\mid D_2, C_1\mid C_2) $. Then, $(D,C)$ is a poly-tiling of $\mathbb{F}_2^{n+m}$ and $D=B_{d_{\max}}({\bzero},r)$.
\end{teo}

\vspace{4pt}
\begin{IEEEproof}
By Corollary \ref{concatenation} we have that $(D,C)$ is a poly-tiling. 
If $\bx\in D$ then $d_{\max}(\bx,{\bzero})=max\{d_{1}(\bx_1,\bzero), d_{2}(\bx_2,\bzero)\}\leq r$, since $\bx_1\in D_1=B_{d_{1}}({\bzero},r)$ and $\bx_2\in D_2=B_{d_{2}}({\bzero},r)$. Thus, $\bx\in B_{d_{\max}}({\bzero},r)$.
 If $\bx=\bx_1\mid \bx_2\notin D$ we have that $\bx_1\notin D_1$ or $\bx_2\notin D_2$, so that $d_{{1}}(\bx_1,\bzero)> r$ or $d_2(\bx_2,\bzero)>r$. But this implies that $d_{\max}(\bx,\bzero)=max\{d_{1}(\bx_1,\bzero),d_{2}(\bx_2,\bzero)\}> r$ and $\bx\notin B_{d_{\max}}({\bzero},r)$. 
Therefore, $D=B_{d_{\max}}({\bzero},r)$.
\end{IEEEproof}

\vspace{3pt}

\begin{exem}
	Let $(D_1,C_1)$ be a tiling of $\mathbb{F}_2^2$, where \[D_1=\{00,10\} \hspace{1cm} C_1=\{00,11\}.\] The poset $P: 1\preceq 2$ can be represented by the Hasse diagram

	\begin{figure}[H]
	\begin{center}	
	\includegraphics[width=0.5cm]{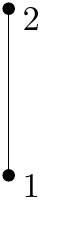}
	\caption{Hasse diagram of the poset $P=([2],\preceq)$}
	\end{center}
	\end{figure}

Notice that, $D_1=B_{d_P}(\bzero, 1)$. That is, $C_1$ is a $(d_P,1)$-perfect code.

	Let $(D_2,C_2)$ be a tiling of $\mathbb{F}_2^4$, with \[D_2=\{0000,1000,0100,0010,0001,1100,1010,1001\}\] $\hspace{0.5cm} C_2=\{0000,1111\}.$ 
	
	\vspace{3pt}
	
	Given $\FF=\{\{1,2\},\{1,3\},\{1,4\}\}$ we have $D_2=B_{d_{\FF}}(\bzero,1)$, which means that $C_2$ is a $(d_{\FF},1)$-perfect code. 
	
	\vspace{3pt}
	
	The concatenation $(D,C)=(D_1\mid D_2,C_1\mid C_2)$ is given by

	\[D=	\left\{
	\begin{array}{ccccccc}
		000000 & \hspace{0.1cm} & 000001 & \hspace{0.1cm} & 100000 & \hspace{0.1cm} & 100001 \\
		001000 & \hspace{0.1cm} & 001100 & \hspace{0.1cm} & 101000 & \hspace{0.1cm} & 101100 \\
		000100 & \hspace{0.1cm} & 001010 & \hspace{0.1cm} & 100100 & \hspace{0.1cm} & 101010 \\		
		000010 & \hspace{0.1cm} & 001001 & \hspace{0.1cm} & 100010 & \hspace{0.1cm} & 101001 \\
	\end{array}
	\right\}   \hspace{1.5cm} C=	\left\{
	\begin{array}{c}
	000000 \\
	001111 \\
    110000 \\
    111111 \\
		\end{array}
	\right\}  \]

\vspace{3pt}

	Notice that $\displaystyle\bigcup_{\bc\in C} \bc+D = \mathbb{F}_2^6$ and $(\bc+D)\cap (\bc'+D)=\emptyset$, for all $\bc,\bc' \in C, \bc\not=\bc'$.
	
	 Moreover, for all $\bx\in D$ we have 
	
	\[d_{\max}(\bx,\bzero)=\begin{cases}
	0, \mbox{if} \ \bx=000000\\
	1, \mbox{if} \ \bx\not= 000000
	\end{cases}\]
	and, $d_{\max}(\bx,\bzero )\geq 2$ for $\bx\notin D$.

It follows that $D=B_{d_{\max}}(\bzero,1)$. In other words, $C$ is a $(d_{\max},1)$-perfect code. 

\end{exem}

\vspace{4pt}
We have just shown in the previous theorem that the concatenation $D=D_1\mid D_2$ of two TS-balls (which are poly-tilings) of \emph{same} radius (possibly determined by different metrics) is a TS-ball. A natural question arises: is it possible to have different radii and $D$ be a ball? 
To answer this question we start constructing a TS-weight, made out of a conditional sum of weights.
 
% If yes, what are the conditions to be considered? Next lemma will provide a metric that turns $D$ a metric ball, considering $D_1$ and $D_2$ as balls with different radius.   

\begin{lema}\label{lema:rs}
		Let $\omega_1$ and $\omega_2$ be TS-weights on $\Fn$ and $\Fm$, respectively, and let $d_i$ be the TS-metric determined by $\omega_i$, $i=1,2$. Given $r\leq n,s\leq m$, let $D_1=B_{d_{1}}({\bzero},r)$,  $D_2=B_{d_{2}}({\bzero},s)$ and $D=D_1\mid D_2$, where $d_i$ is the metric determined by $\omega_i$. We define the $(r,s)$-sum

\[\omega_1 \oplus_{s}^r   \omega_2(\bx) = 
\begin{cases}
\omega_1(\bx_1) +\omega_2(\bx_2), & \mbox{ if }  \bx\in D\\
r+s+1, & \mbox{ otherwise.} 
\end{cases}
\]
Then, $ \omega_1 \oplus_{s}^r   \omega_2 $ is a weight and it respects support.
\end{lema}

\begin{IEEEproof}

\begin{enumerate}
	\item [i)] Given $\bx\in \Fmn$. If $\bx\in D$, then $\omega_1 \oplus_{s}^r   \omega_2(\bx)=\omega_1(\bx_1)+\omega_2(\bx_2)\geq 0$. Otherwise, $\omega_1 \oplus_{s}^r   \omega_2(\bx)=r+s+1>0$, which implies that $\omega_1 \oplus_{s}^r   \omega_2(\bx)\geq 0$, for every $\bx\in\Fmn$. 
	\item [ii)] Let $\bx=\bzero$. Since $\bzero \in D$, then $\omega_1 \oplus_{s}^r   \omega_2(\bx)=\omega_1(\bzero_m)+\omega_2(\bzero_n)=0$. Conversely, if $\omega_1 \oplus_{s}^r   \omega_2(\bx)=0$ then $\bx\in D$ and $\omega_1(\bx_1)+\omega_2(\bx_2)=0$. Since $\omega_1$ and $\omega_2$ are weights, we have $\bx=\bzero$. 
	\item [iii)] Given $\bx,\by\in\Fmn$. If $\bx+\by\in D$ then $\omega_1 \oplus_{s}^r   \omega_2(\bx+\by)=\omega_1(\bx_1+\by_1)+\omega_2(\bx_2+\by_2)\leq \omega_1(\bx_1)+\omega_1(\by_1)+\omega_2(\bx_2)+\omega_2(\by_2)=\omega_1(\bx_1)+\omega_2(\bx_2)+\omega_1(\by_1)+\omega_2(\by_2)=\omega_1 \oplus_{s}^r   \omega_2(\bx)+\omega_1 \oplus_{s}^r   \omega_2(\by)$.
	
Now, suppose $x+y\notin D$. Then $\omega_1 \oplus_{s}^r   \omega_2(\bx+\by)=r+s+1\leq 2r+2s+2=\omega_1 \oplus_{s}^r   \omega_2(\bx) + \omega_1 \oplus_{s}^r   \omega_2(\by)$.
\end{enumerate}

To complete the proof, it remains to show that $\omega_1 \oplus_{s}^r   \omega_2$ respects support. 

Let $\bx,\by\in \Fmn$ such that $\supp(\bx)\subseteq \supp(\by)$. If $\bx,\by\in D$ then $\omega_1 \oplus_{s}^r   \omega_2(\bx)=\omega_1(\bx_1)+\omega_2(\bx_2)$ and $\omega_1 \oplus_{s}^r   \omega_2(\by)=\omega_1(\by_1)+\omega_2(\by_2)$. Since $\omega_1(\bx_1)\leq \omega_1(\by_1)$ and $\omega_2(\bx_2)\leq \omega_2(\by_2)$, we have that $\omega_1 \oplus_{s}^r \omega_2(\bx)=\omega_1 \oplus_{s}^r \omega_2(\by)$. 

If $\bx\in D$ and $\by\notin D$, then $\omega_1 \oplus_{s}^r \omega_2(\bx)=\omega_1(\bx_1)+\omega_2(\bx_2)\leq r+s< r+s+1=\omega_1 \oplus_{s}^r \omega_2(\by)$.

Finally, if $\bx,\by\notin D$ then $\omega_1 \oplus_{s}^r \omega_2(\bx)=\omega_1 \oplus_{s}^r \omega_2(\by)=r+s+1$. 

Therefore, $\omega_1 \oplus_{s}^r \omega_2$ is a weight which respects the support of vectors.
	\end{IEEEproof}
\vspace{3pt}

\begin{teo}\label{teo:tilingposetconcat}
Let $C_1,C_2$ be TS-perfect codes. Then, $C=C_1\mid C_2$ is a TS-perfect code.
\end{teo}

\begin{IEEEproof}
	Given that $C_1,C_2$ are TS-perfect codes, there are $d_1\in\mathcal{TS}(n), d_2\in\mathcal{TS}(m)$ such that the balls  $D_1=B_{d_{1}}({\bzero},r)$ and $D_2=B_{d_{2}}({\bzero},s)$, for some $r,s$ are such that $(D_1,C_1)$ and $(D_2,C_2)$ are convex poly-tilings of $\Fn$ and $\Fm$ respectively. We denote by $\omega_1,\omega_2$ the weights determined by $d_1,d_2$ respectively.
	
	Let $(D,C) =(D_1\mid D_2,C_1 \mid C_2)$ be the concatenated tiling. 
 Corollary \ref{concatenation} ensures that $(D,C)$ is a poly-tiling of $\Fnm$.  Let  $\omega_r^s:=\omega_1 \oplus_s^r   \omega_2$ be defined as in Lemma \ref{lema:rs}. The definition of $\omega_r^s$ ensures that $D=\{ \bx\in\Fnm ;  \omega_r^s(\bx) \leq r+s  \}    $. The lemma ensures that  $D$ is the ball of radius $r+s$ of the metric determined by $\omega_r^s$. Since this is a TS-metric, Proposition \ref{prop:poly} ensures that $C$ is a $(d_r^s,r+s)$-perfect code.
  \end{IEEEproof}

%\textcolor{blue}{---------------- Seção nova ADICIONADA ----------------}

\subsection{Concatenation of balls: the case of combinatorial metrics}

In the previous section, it was proved that, given two balls of two TS-metrics, its concatenation is also a metric ball of a third TS-metric, obtained by a conditional sum of the given ones.
In this section, we prove that if the original metrics are combinatorial ones, also the former metric (used to turn the concatenated code to be perfect) can be taken as a combinatorial metric. %concatenation of two $d_{\mathcal{F}}$-perfect codes with same radius results in a $d_{\mathcal{F}}$-perfect code, that is, concatenation of combinatorial perfect codes results in a combinatorial perfect code. The proof consists in to find an appropriated combinatorial metric.    

\vspace{3pt}

\begin{prop}\label{ladrilhocombin}
	Let $(D_1,C_1)$ and $(D_2,C_2)$ be poly-tilings of $\mathbb{F}_2^n$ and $\mathbb{F}_2^m$, respectively. Suppose that $D_1=B_{d_{\mathcal{F}_1}}({\bzero},r)$ and $D_2=B_{d_{\mathcal{F}_2}}({\bzero},r)$. Then, the concatenation $(D,C):=(D_1\mid D_2,C_1\mid C_2)$ is a poly-tiling of $\mathbb{F}_2^{n+m}$ and $D=B_{d_{\mathcal{F}}}({\bzero},r)$, where $\mathcal{F}=\mathcal{F}_1\ast \mathcal{F}_2:=\{F_1\cup(n+F_2), F_1\in \mathcal{F}_1,F_2\in \mathcal{F}_2    \}$.
\end{prop}

\begin{IEEEproof}
	First of all, we note that, for $F_2\in\FF_2$, we have that $n+F_2:=\{i+n; i\in F_2  \}\subset [n+m]\setminus [n]$, and since $F_2$ is a covering of $[m]$, we have that $n+F_2$ is a covering of $[n+m]\setminus [n]$.
	
	By Corollary \ref{concatenation}, we have $(D,C)$ is a poly-tiling of $\Fnm$ and we need to prove that  $D=B_{d_{\mathcal{F}}}({\bzero},r)$.
	
	Given $\bx\in D$ let us write $\bx=\bx_{1}\mid \bx_{2}$, where $\bx_{i}\in D_i, i =1,2$. Notice that $\omega_{\mathcal{F}_1}(\bx_1)\leq r$ and $\omega_{\mathcal{F}_2}(\bx_2)\leq r$, since $\bx_1\in D_1, \bx_2\in D_2$. In other words, there exists $\mathcal{A}_1\subset\mathcal{F}_1$ such that $|\mathcal{A}_1|\leq r$ and $\mathcal{A}_1$ is a covering of $\supp(\bx_1)$ and there is $\mathcal{A}_2\subset\mathcal{F}_2$ such that $|\mathcal{A}_2|\leq r$ and $\mathcal{A}_2$ is a covering of $\supp(\bx_2)$. Let us write $\mathcal{A}_i=\{ F_{i,1},F_{i,2}, \ldots , F_{i,r} \}$ for $i=1,2$, possibly having some of the $F_{i,j}=\emptyset$ in case $|\mathcal{A}_i|<r$. There is some abuse of notation in here, since we are admitting the possibility that more than one element of the family to be the empty set, but this shall cause no harm. 
	Now, we notice that 	$\supp(\bx)=\supp(\bx_1)\cup (\supp(\bx_2)+n)$ and it follows that 
	\[
	\supp(\bx) \subset \left[F_{1,1}\cup (n+F_{2,1})\right]\cup \ldots \cup \left[F_{1,r}\cup( n+F_{2,r})\right],
	\]
%	
%	hence $\bx\in B_{\mathcal{F}}(\bzero,r)$.
%	
%	
	that is, it is possible to use $r$ elements of  $\FF$ to cover $\supp(\bx)$, thus $\bx\in B_{d_{\FF}}({\bzero},r)$.
	
	Let $\by\in B_{d_{\FF}}({\bzero},r)$ and write $\by=\by_1\mid\by_2$. Since $\by\in B_{d_{\FF}}({\bzero},r)$, there are $F_1,F_2,\ldots ,F_r \in \mathcal{F}$ such that 
	\[
	\supp (\by)\subset \bigcup_{i=1}^rF_i.
	\] 
	For each $i\leq r$ we can write $F_i=F_{i,1}\cup (n+F_{i,2})$ where $F_{i,j}\in\mathcal{F}_i$, for $j=1,2$. But $\supp(\by)$ can be expressed as the disjoint union  $\supp(\by_1)\cup (n+\supp (\by_2))$ and $\supp (\by)\subset \cup_{i=1}^rF_i$ implies that $\supp (\by_1\mid\bzero )\subset \bigcup_{i=1}^rF_{i,1}$ and $\supp (\bzero\mid\by_2)\subset \bigcup_{i=1}^r(n+F_{i,2})$. It follows that  $\by_1\in B_{\mathcal{F}_1}(\bzero ,r)$ and  $\by_2\in B_{\mathcal{F}_2}(\bzero ,r)$. 
\end{IEEEproof}

%	But
%	\begin{align*}
%\supp (\by_1) &\subset (\cup_{i=1}^rF_i)\cap [n]\\
%&=\cup_{i=1}^r(F_i \cap [n])\\
%&=\cup_{i=1}^rF_{1,i},
%\end{align*}	
%
%hence $\by_1\in B_{\mathcal{F}_1}(\bzero ,r)$ and in a similar way we get that $\by_2\in B_{\mathcal{F}_2}(\bzero ,r)$.
%	
%	-------------------------
	
\vspace{3pt}
	
%	Notice that is possible to write $\by=\by_1\mid \by_2, \by_1\in \Fn, \by_2\in \Fm$. Suppose that $\by\notin D$, then we have two possibilities:  $\by_1\notin D_1$ or $\by_2\notin D_2$. %2) $\by_1\notin D_1$ and $\by_2\notin D_2$. 
%	
%	In the first case, suppose, wlog, that $\by_1\notin D_1$, then there exists $F_1\subset \FF$ where $F_1$ is a covering of $\by_1$ and $|F_1|> r$. Since $\FF=\FF_1\ast \FF_2$, there exists $F\subset \FF$ such that 
%	$$F=F_1\cup (F_2+n)$$
%	where $F_2$ is a covering of $\supp(\by_2)$.
%	
%	Thus, $F$ is a covering of $\supp(\bx)$ and $|F|>r$. This is a contradiction, since $\by\in B_{{d_{\FF}}}({\bzero},r)$. Then, $\by_1\in D_1$ which implies that $\by\in D$. 
%	
%	
%	In the case where $\by_1\notin D_1$ and $\by_2\notin D_2$, there exist $F_1$ covering of $\supp(\by_1)$ and $F_2$ covering of $\supp(\by_2)$ such that $|F_1|> r$ and $|F_2|> r$, then there exists $F\subset \FF$ such that $F=F_1\cup (F_2+n)$, $|F|>r$, which is a contradiction, since $\by\in B_{d_{\FF}}({\bzero},r)$.
%	
%	Therefore, $D=B_{d_{\FF}}({\bzero},r)$.   
%

 Proposition \ref{ladrilhocombin} demands that both the original perfect codes  to have the same radius. But what happens when the radii are different? In this case, an alternative is to define a new covering that reduces any ball into a ball with radius 1. 

\begin{defi}\label{cobertura}
	Let $\FF$ be a covering of $[n]$ and $D\subseteq \Fn$ be a ball, $D=B_{d_{\FF}}({\bzero},r)$. The $(D,\FF)$\textbf{-covering} $\FF_D$ of $[n]$ is defined as $$\FF (D)=\FF\cup \left\{ \supp(\bx);\bx\in D \right\}.$$% where $\bx\in D$ and $\supp(\bx)\not\subseteq \FF$.
\end{defi}

\vspace{3pt}

%The idea of the next proposition is to define another covering $\mathcal{F}^{\ast}$ (given in Definition \ref{cobertura}) and to show that $D$ is a $d_{\FF^{\ast}}$-ball with radius 1. The case where $D_1$ and $D_2$ has different radius can be solved by applying Proposition \ref{festrela} to both $D_1$ and $D_2$, resulting in balls of radius 1 and then we may apply Proposition \ref{ladrilhocombin} obtaining that $D$ is a ball. 
It is pretty intuitive that $\FF (D)$ turns $D$ into a ball of radius $1$. Nevertheless, we prove it in the next lemma.
\vspace{3pt}

\begin{lema}\label{festrela}
	Let $D=B_{d_\FF}(\bzero,r)$, with $r$ positive. Then, % the covering $\FF_D$ turns $D$ into a ball of radius $1$, that is,  
	$D=B_{d_{\FF (D)}}(\bzero,1)$.  
\end{lema}

\begin{IEEEproof}
	Given $\bzero\neq\bx\in D$, it is clear that $\omega_{\FF (D) }(\bx)=1$, since  $\supp(\bx)\in \FF (D)$, hence $D\subset B_{d_{\FF (D)}}(\bzero,1)$. 
	
	Let us now consider $\bx\in B_{d_{\FF (D)}}(\bzero,1)$. Then, for $\bx\neq\bzero$, there is $F\in\FF (D)$ such that $\supp (\bx)\subset F$. If $F\in\FF$ we have that $\bx\in B_\FF (\bzero ,1)\subset B_\FF (\bzero , r)=D$. If this is not the case, there is $\by\in D$ such that $\supp (\bx)\subset \supp (\by)$. But the metric $d_\FF$ respects the support and since $\by\in B_{d_\FF}(\bzero ,r) $ and $\supp (\bx)\subset \supp (\by)$, we have that also $\bx\in B_{d_\FF}(\bzero ,r)=D $, that is,   $ B_{d_{\FF (D)}}(\bzero,1)\subset D$.
%	
%	
%	Suppose now that $\by\notin D$. Let $F_1,F_2,\ldots F_s\in\FF$ be a minimal family of sets such that $\supp (\by )\subset \cup_{i=1}^sF_i$. Since $D=B_{d_\FF}(\bzero,r)$ we have that $s>r$. 
%	
%	 Suppose that there exists $\bx\in D$ such that $\supp(\bx)\not\subset \FF$. Let $\supp(\bx)=\{1,2,\ldots,s\}$, $s\leq n$. Then $\FF^{\ast}=\FF \cup \{1,2,\ldots,s\}$.  
%	
%	Notice that, since $D=B_{d_{\FF}}({\bzero},r)$ and $D$ respects support, then for every $\by\in \Fn$ such that $\supp(\by)\subset \supp(\bx)$, we have $\by\in D$. Thus, for every $\bz\in \Fn$ such that $\supp(\bz)\subset \{1,2,\ldots,s\}$, we have $d_{\FF^{\ast}}({\bzero},\bz)=1$. 
%	
%	Let $\bx_1\in D$ such that $d_{\FF^{\ast}}({0},\bx_1)\geq 2$. We have $d_{\FF}({\bzero},\bz)\leq r$, since $D$ is a $d_{\FF}$-ball with radius $r$, then there exists $\bx_2\in D$ such that $\supp(\bx_1)\subset \supp(\bx_2)$ and $d_{\FF^{\ast}}({\bzero},\bx_2)=1$. 
%	
%	For each $\bx_i\in D$, $|\supp(\bx_i)|\leq 2$, there exists $\bx_j\in D$ such that $\supp(\bx_i)\subset \supp(\bx_j)$, where $\supp(\bx_j)\not\subset \FF$, then we add $\supp(\bx_j)$ to $\FF^{\ast}$. Thus, for every $\bx\in D$, we have $d_{\FF^{\ast}}({\bzero},\bx)=1$.
%	
%	Therefore, $D=B_{d_{\FF^{\ast}}}({\bzero},1)$.
\end{IEEEproof}

\vspace{3pt}

\begin{prop}
	If $D_i$ is a $d_{\FF_i}$-ball, $i=1,2$ then $D=D_1\mid D_2$ is a $d_{\FF}$-ball, where $\FF=\FF_1 (D_1)  \ast\FF_2 (D_2)$.
\end{prop}

\begin{IEEEproof}
	By Lemma \ref{festrela} we have that if $D_1=B_{d_{\FF_1}}({\bzero},r_1)$ and $D_2=B_{d_{\FF_2}}({\bzero},r_2)$ then $D_1=B_{d_{\FF_1 (D_1)}}({\bzero},1)$ and $D_2=B_{d_{\FF_2 (D_2)}}({\bzero},1)$. By Proposition \ref{ladrilhocombin} we have $D=D_1\mid D_2$ is a $d_{\FF }$-ball with radius 1, that is, $D=B_{d_{\FF }}({\bzero},1)$.    
\end{IEEEproof}

\vspace{3pt}
The tool used in Definition \ref{cobertura} is a way to solve the concatenation problem when two balls have distinct radius, because Proposition \ref{ladrilhocombin} is applicable only to the case when the balls have the  same radius. This is a somehow artificial construction and we tried to make it better by reducing the largest radius in order to make it fits the smaller.  However, this is not always possible  if the smaller radius is greater than $1$, as we can see in the next example. 

\begin{exem}
Let $\FF=\{\{1\},\{2\},\{3\},\{4\},\{5\},\{6\}\}$ be a covering of $[6]$, which defines the Hamming metric on $\mathbb{F}_2^6$.  Suppose that $\FF'$ is a cover of $[6]$ such that $B_{\FF}(\bzero,3)=B_{\FF'}(\bzero,2)$. 

Let us consider a vector $\bx \in \mathbb{F}_2^6$ such that $\omega_\FF (\bx)=3$ and $\omega_{\FF'}(\bx)=2$. This implies the existence of a vector $\bx_1$ with $\supp (\bx_1)\subset \supp (\bx)$, $\omega_\FF(\bx_1)=2$ and $\omega_{\FF'}(\bx_1)=1$.

  The vector $\by:=\bx +111111$ also has Hamming weight (the $\FF$-weight) equals to $3$. Assuming that $B_{d_\FF}(\bzero ,3)=B_{d_{\FF'}}(\bzero ,2) $ we get that there is a vector $\by_1$ with $\supp (\by_1)\subset \supp (\by)$, $\omega_{\FF'}(\by_1)=1$ and  $\omega_\FF(\by_1)=2$   or $\omega_\FF(\by_1)=3$.
  
  Let $\bz:=\bx_1+\by_1$. We have that $\omega_{\FF'}(\bz)=2$ but $\omega_\FF (\bz)$ equals $4$ or $5$, according to the $\FF$-weight of $\by_1$ being equal to $2$ or $3$.
  
%  -----------------------  
%  either $\omega_{\FF'}(\by)=2$ or $\omega_{\FF'}(\by)=3$. In the first case we find
%
% Since both the metri the vectors $\bx=111000,\by=000111$ with Hamming weight   $\supp(111000),\supp(000111)$
%
%But, notice that to reduce the radius of the ball, all the vectors $\bx \in B_{\FF}(\bzero,3)$ that have weight equal to 3 need to have weight equal to 2 when considering $\FF'$. To do so, there will be $A\subset \FF'$ with $|A|=2$ and $A$ a covering of the $\supp(\bx)$. Then, to cover all the vectors with weight 3, $A_i\cap A_j\not= \emptyset$, $A_i, A_j$ coverings of $\supp(\bx_i),\supp(\bx_j)$, otherwise, the covering $\FF'$ will generate other vectors that not belong to $B_{\FF}(\bzero,3)$. But, the coverings of $\supp(111000),\supp(000111)$ have no intersection. Then, it is not always possible to reduce the radius of $B_{\FF}(\bzero,3)$.    
%
%
%Then both the vectors $111000$ and $000111$  belong to $B_{\FF}(\bzero,3)$.

%\[B_{\FF}(\bzero,3)= \left\{
%\begin{array}{ccccc}
%000000 & 110000 & 010010 & 000011 & 101001   \\
%100000 & 101000 & 010001 & 111000 & 100110 \\
%010000 & 100100 & 001100 & 110100 & 100101 \\
%001000 & 100010 & 001010 & 110010 & 100011\\
%000100 & 100001 & 001001 & 110001 & 011100  \\
%000010 & 011000 & 000110 & 101100 & 010110 \\
%000001 & 010100 & 000101 & 101010 & 010101 \\
%010011 & 001110 & 001101 & 001011 & 000111 \\
%\end{array}
%\right\}.\]	
\end{exem}

\section*{Acknowledgment}
Gabriella Akemi Miyamoto was supported by Capes (finance code 001) and CNPq. Marcelo Firer was partially supported by Sao Paulo Research Foundation, (FAPESP grant 2013/25977-7) and CNPq.

\bibliography{ref.bbl}{}

\begin{thebibliography}{10}

\bibitem{alon}
N.~Alon and R.~Yuster.
\newblock {\em H-factors in dense graphs}.
\newblock J. Combin. Theory Ser. B 66, pp. 269–282, n. 2, 1996.

\bibitem{muniz}
M.~M.~S. Alves, L.~Panek, and M.~Firer.
\newblock {\em Error-block codes and poset metrics}.
\newblock Adv. Math. Commun., pp. 95, vol. 2, 2008.

\bibitem{balogh}
J.~Balogh, A.~Treglown, and A.~Z. Wagner.
\newblock {\em Tilings in Randomly Perturbed Dense Graphs}.
\newblock Combinatorics, Probability and Computing, Cambridge University Press,
  vol. 28, pp. 159–176, 2019.

\bibitem{bossert}
M.~Bossert and V.~Sidorenko.
\newblock {\em Singleton-type bounds for blot-correcting codes}.
\newblock IEEE Trans. Inf. Theory 42 (3), pp. 1021–1023, 1996.

\bibitem{branko}
G.~Branko and G.~C. Shephard.
\newblock {\em Tilings and Patterns}.
\newblock W. H. Freeman \& Co., 1986.

\bibitem{brualdi}
R.~A. Brualdi, S.~Graves, and K.~M. Lawrence.
\newblock {\em Codes with a poset metric}.
\newblock Discrete Mathematics, pp. 57–72, v. 147, 1995.

\bibitem{vardy}
G.~Cohen, S.~Litsyn, A.~Vardy, and G.~Zémor.
\newblock {\em Tilings of binary spaces}.
\newblock SIAM J. Discrete Math. 9-3, pp. 393-412, 1996.

\bibitem{dass}
B.K. Dass, N.~Sharma, and R.~Verma.
\newblock {\em Perfect codes in poset spaces and poset block spaces}.
\newblock Finite Fields Appl., pp. 90-106, vol. 46, 2017.

\bibitem{deza2009}
M.~M. Deza and E.~Deza.
\newblock {\em Encyclopedia of distances}.
\newblock Springer, pp. 1-583, 2009.

\bibitem{rafael}
R.~G.~L. D'Oliveira and M.~Firer.
\newblock {\em Geometry of communication channels: metrization and decoding}.
\newblock Symmetry Culture And Science, pp. 279-289, v. 27, 2016.

\bibitem{rafael2}
R.~G.~L. D'Oliveira and M.~Firer.
\newblock {\em Channel metrization}.
\newblock European Journal of Combinatorics,
  https://doi.org/10.1016/j.ejc.2018.02.026., 2018.

\bibitem{feng}
K.~Feng, L.~Xu, and F.J. Hickernell.
\newblock {\em Linear error-block codes}.
\newblock Finite Fields Appl. 12 (4), pp. 638–652, 2006.

\bibitem{marcelo}
M.~Firer, M.M.S. Alves, J.A. Pinheiro, and L.~Panek.
\newblock {\em Poset Codes: partial orders, metrics and coding theory}.
\newblock Springer Briefs in Mathematics, Springer, 2018.

\bibitem{gruslys2}
Vytautas. G.
\newblock {\em Decomposing the vertex set of a hypercube into isomorphic
  subgraphs}.
\newblock arXiv:1611.02021, 2016.

\bibitem{gabidulin}
E.~M. Gabidulin.
\newblock {\em Combinatorial metrics in coding theory}.
\newblock 2nd International Symposium on Information Theory, 1973.

\bibitem{golay1}
M.~J.~E. Golay.
\newblock {\em Notes on digital coding}.
\newblock Proc. IEE 37, pp. 657, 1949.

\bibitem{golomb2}
S.~W. Golomb and L.~R. Welch.
\newblock {\em Algebraic coding and the Lee metric}.
\newblock Error Correcting Codes, Wiley, New York, pp. 175-189, 1968.

\bibitem{golomb}
S.~W. Golomb and L.~R. Welch.
\newblock {\em Perfect codes in the Lee metric and the packing of polyominoes}.
\newblock SIAM J. Appl. Math., 18 (2), pp. 302–317, 1970.

\bibitem{gruslys}
V.~Gruslys, I.~Leader, and I.~Tomon.
\newblock {\em Partitioning the Boolean lattice into copies of a poset}.
\newblock Journal of Combinatorial Theory, Series A, vol. 161, pp. 81-98, 2019.

\bibitem{hajnal}
A.~Hajnal and E.~Szemeredi.
\newblock {\em Proof of a conjecture of P. Erdos}.
\newblock 1970.

\bibitem{hamming}
R.~W. Hamming.
\newblock {\em Error-detecting and error-correcting codes}.
\newblock Bell System Tech. J. 29, pp. 147-160, 1950.

\bibitem{olof}
O.~Heden.
\newblock {\em A survey of perfect codes}.
\newblock Adv. Math. Commun. pp. 223-247, vol. 2, 2008.

\bibitem{hyun}
J.Y. Hyun and H.K. Kim.
\newblock {\em The poset structures admitting the extended binary Hamming code
  to be a perfect code}.
\newblock Discrete Mathematics, pp. 37-47, vol. 288, 2004.

\bibitem{jang}
C.~Jang, H.K. Kim, D.Y. Oh, and Y.~Rho.
\newblock {\em The poset structures admitting the extended binary Golay code to
  be a perfect code}.
\newblock Discrete Mathematics, pp. 4057-4068, n. 18, vol. 308, 2008.

\bibitem{jang2}
Y.~Jang and J.~Park.
\newblock {\em On a MacWilliams type identity and a perfectness for a binary
  linear (n,n-1,j) - poset code}.
\newblock Discrete Mathematics, pp. 85-104, n. 1, vol. 265, 2003.

\bibitem{kim}
H.~K. Kim and D.~Y. Oh.
\newblock {\em On the nonexistence of triple-error-correcting perfect binary
  linear codes with a crown poset structure}.
\newblock Discrete Mathematics, pp. 174 - 181, n. 1, vol. 297, 2005.

\bibitem{kuhn}
D.~Kuhn and D.~Osthus.
\newblock {\em The minimum degree threshold for perfect graph packings}.
\newblock Combinatorica 29, pp. 65–107, n. 1, 2009.

\bibitem{lint}
J.H.~Van. Lint.
\newblock {\em A survey of perfect codes}.
\newblock Rocky Mountain J. Math. 5, pp. 199-224, n. 2, 1975.

\bibitem{roberto}
R.~A. Machado.
\newblock {\em Weights which respect support and NN-decoding}.
\newblock arXiv:1804.07809, 2018.

\bibitem{akemi}
G.~A. Miyamoto.
\newblock {\em Equivalence classes of small tilings of the Hamming cube}.
\newblock arXiv:1904.11034, 2019.

\bibitem{panek}
Firer~M. Panek, L. and M.~M.~S. Alves.
\newblock {\em Classification of Niederreiter–Rosenbloom–Tsfasman Block
  Codes}.
\newblock IEEE Transactions on Information Theory, pp. 5207-5216, n. 10, vol.
  56, 2010.

\bibitem{jerry}
J.A. Pinheiro, R.A. Machado, and M.~Firer.
\newblock {\em Combinatorial metrics: MacWilliams-type identities, isometries
  and extension property}.
\newblock Designs, Codes and Cryptography, Issue 2–3, pp. 327–340, vol. 87,
  2019.

\bibitem{post}
K.~A. Post.
\newblock {\em Nonexistence theorem on perfect Lee codes over large alphabets}.
\newblock Information and Control 29, pp. 369–380, 1975.

\end{thebibliography}
\bibliographystyle{plain}

\end{document}